\documentclass{aa}  

\usepackage[varg]{txfonts}
\usepackage{rotating}
\usepackage{ulem}
\usepackage[breaklinks, colorlinks, citecolor=blue, urlcolor = blue]{hyperref}
      
\usepackage{csquotes}
\usepackage{units}
\usepackage{mathtools}
\usepackage{float}
\usepackage[caption = false]{subfig}
\usepackage{graphicx}
\usepackage{amsmath}
\usepackage{footmisc}

\bibpunct{(}{)}{;}{a}{}{,} 

\newcommand \fe{{\it Fermi}}
\newcommand \eb{{$E_{\rm break}$}}
\newcommand \ep{{$E_{\rm peak}$}}
\newcommand{\order}[1]{} 

\begin{document} 

\title{Consistency with synchrotron emission in the bright GRB~160625B observed by {\it Fermi}}
\titlerunning{Synchrotron emission in GRB~160625B}
\author{M.~E. Ravasio\inst{\ref{inst1},\ref{inst2}}
\and G. Oganesyan\inst{\ref{inst3}}
\and G. Ghirlanda\inst{\ref{inst2},\ref{inst4}}
\and L. Nava\inst{\ref{inst2},\ref{inst5},\ref{inst6}}
\and G. Ghisellini\inst{\ref{inst2}}
\and A. Pescalli\inst{\ref{inst2},\ref{inst7}}
\and A. Celotti\inst{\ref{inst2},\ref{inst3},\ref{inst6}}
}

\institute{ Universit\`a degli Studi di Milano-Bicocca, Dipartimento di Fisica U2, Piazza della Scienza, 3, I--20126, Milano, Italy \\
\email{\href{m.ravasio5@campus.unimib.it}{m.ravasio5@campus.unimib.it}}\label{inst1}
\and
INAF -- Osservatorio Astronomico di Brera, via Bianchi 46, I--23807 Merate (LC), Italy \label{inst2}
\and
SISSA, via Bonomea 265, I--34136 Trieste, Italy \label{inst3}
\and
INFN -- Milano Bicocca, Piazza della Scienza 3, I--20123, Milano, Italy \label{inst4}
\and
INAF -- Osservatorio Astronomico di Trieste, via G.B. Tiepolo 11, I--34143 Trieste, Italy \label{inst5}
\and
INFN -- via Valerio 2, I-34127 Trieste, Italy \label{inst6}
\and
Universit\`a degli Studi dell’Insubria, Via Valleggio, 11, I-22100 Como, Italy\label{inst7}
}  

\abstract{
We present time-resolved spectral analysis of prompt emission from GRB~160625B, one of the brightest bursts ever detected by {\it Fermi} in its nine years of operations. Standard empirical functions fail to provide an acceptable fit to the GBM spectral data, which instead require the addition of a low-energy  break to the fitting function. 
We introduce a new fitting function, called 2SBPL, consisting of three smoothly connected power laws. Fitting this model to the data, the goodness of the fits significantly improves and the spectral parameters are well constrained. 
We also test a spectral model that combines non-thermal and thermal (black body) components, but find that the 2SBPL model is systematically favoured.
The spectral evolution shows that the spectral break is located around $E_{\rm break}\sim$100\,keV, while the usual $\nu F_{\nu}$ peak energy feature $E_{\rm peak}$ evolves in the $0.5-6$\,MeV energy range.
The slopes below and above $E_{\rm break}$ are consistent with the values --0.67 and --1.5, respectively, expected from synchrotron emission produced by a relativistic electron population with a low-energy cut-off. 
If $E_{\rm break}$ is interpreted as the synchrotron cooling frequency, the implied magnetic field in the emitting region is $\sim$ 10 Gauss, i.e. orders of magnitudes smaller than the value expected for a dissipation region located at $\sim 10^{13-14}$\,cm from the central engine.
The low ratio between $E_{\rm peak}$ and $E_{\rm break}$ implies that the radiative cooling is incomplete, contrary to what is expected in strongly magnetized and compact emitting regions. 
}


\maketitle

\section{Introduction}\label{sec:intro}
The physics of gamma-ray burst (GRB) prompt emission is still debated. 
The main radiative process responsible for the observed soft $\gamma$-ray emission has not been clearly identified even fifty years after their discovery. 
Independently from the details of the acceleration mechanisms (shocks at internal collisions or in magnetic reconnection events), the presence of energized electrons and strong magnetic fields points toward synchrotron radiation as the most natural and efficient process for conversion of particle energy into non-thermal $\gamma$-ray radiation \citep{Rees94,Katz94,Tavani96,Sari96,Sari98}. 
The simplest predictions from synchrotron radiation in the fast cooling regime are, however, inconsistent with the shape of the observed prompt emission spectra.
Below the peak energy of the $\nu F_{\nu}$ spectrum, the photon index  indeed has a typical value $\langle\alpha\rangle \sim -1$ \citep{Preece1998,Frontera2000,Ghirlanda2002,Kaneko2006,Sakamoto2011,Nava2011,Goldstein12,Gruber2014,Lien2016}.
This slope is harder than the predicted $\alpha^{\rm syn}=-1.5$ expected in the case of synchrotron emission from a population of non-thermal electrons undergoing efficient cooling \citep{Ghisellini2000}. 

Different solutions for this major inconsistency have been proposed in the literature. 
Within the synchrotron scenario, harder spectra can be achieved in some non-standard configurations.
These include a magnetic field decaying downstream on timescales  shorter than the dynamical one \citep{Peer2006,Zhao2014} or decaying with the distance from the central engine \citep{Uhm2014}, highly anisotropic magnetic fields \citep{Medvedev2000},  synchrotron self-absorption frequency close to the X-ray range \citep{Lloyd2000,Daigne11}, inverse Compton scattering in the Klein--Nishina regime \citep{Derishev01,Nakar09,Daigne11}. 

In two recent studies, \cite{gor2017a} and \cite{gor2017b} extended the investigation of the prompt emission spectra down to the soft X-ray band, taking advantage of 34 {\it Swift} GRBs with prompt emission observed not only by the  Burst and Alert Telescope (BAT; 15--150\,keV), but also by the X-ray Telescope (XRT; 0.3--10\,keV). 
Time-resolved XRT+BAT joint spectral analysis has revealed the necessity of going beyond the standard fitting models, by adding a further, hard power-law segment at low energy.
In the GRBs considered by \cite{gor2017a,gor2017b}, the energy $E_{\rm break}$ at which the spectrum breaks, assumes values in the range 2--30\,keV.
In both studies, this break is required with high statistical significance (more than 3$\sigma$) in $\sim$\,65\% of the analysed spectra.
These spectra also display a peak energy $E_{\rm peak}$, with values similar to those of the whole population, i.e. ranging from 10\,keV to 1\,MeV.
The photon index $\alpha_1$ describing the spectrum below the break energy has a distribution peaked around --0.6, while the photon index $\alpha_2$ describing the spectral segment between \eb\ and \ep\ has a mean value of --1.5.
The similarity between these mean values and the values expected from synchrotron fast cooling radiation ($\alpha_1^{\rm syn}=-0.67$ and $\alpha_2^{\rm syn}=-1.5$) lead to identifying \eb\ with the cooling frequency and \ep\ with the characteristic synchrotron frequency.
Moreover, the relatively low ratio $E_{\rm peak}/E_{\rm break} \sim 30$ found in these spectra corresponds to a regime of moderately fast cooling. 

Since the \eb\ distribution inferred by these first studies extends up to 30\,keV, an immediate follow-up question is whether in the \fe\ database there are bursts showing a similar spectral break to those found by \cite{gor2017a,gor2017b}. 
Thanks to the energy range of sensitivity characterizing the Gamma Burst Monitor  (GBM;  8\,keV--40\,MeV) on board \fe, it should be possible to find these breaks and study whether they are also present at higher energies.
Moreover, if a low-energy spectral break is found in bright GBM GRBs, a time-resolved analysis would then allow to study, for the first time in detail, if and how this break energy evolves in time and with respect to the peak energy.

In this work, we consider a test case event (GRB~160625B) satisfying two conditions: it has a great deal of  photon statistics 
 and it is poorly fitted by standard fitting functions, as reported in the online GBM GRB spectral catalogue\footnote{https://heasarc.gsfc.nasa.gov/W3Browse/fermi/fermigbrst.html\label{fn:GBMcatalog}}.    

According to the GBM Catalog, GRB~160625B is the third burst with the largest fluence 
(5.7$\times 10^{-4}$\,erg\,cm$^{-2}$ in the $10-10^3$\,keV energy range) detected by \fe. 
At a redshift of $z=1.406$ \citep{xu16} its isotropic energy is $E_{\rm iso} \sim 5 \times 10^{54}$\,erg. 
This GRB has been extensively studied in the literature, due to its extremely large fluence and long duration \citep{Zhang2016,Wang2017,Lu2017}, to the rich data sets covering its afterglow emission, and to polarization measurements \citep{Alexander2017,Troja2017}.

Time-resolved prompt spectral analysis performed by \cite{Zhang2016}, suggested the presence of a black-body (BB) spectrum in the first peak (the precursor), and a non-thermal spectrum during the main emission episode. This spectral transition was interpreted as being caused by the transition from a matter-dominated jet to a magnetically dominated jet. 
\cite{Wang2017} adopted a composition of Band function \citep{Band1993} with a high-energy cut-off and BB component. 
A similar two-component model is adopted by \cite{Lu2017}. 
What appears common in these models is the presence, sometimes simultaneous, of a BB and a non-thermal component. 

We revisit these analyses in light of the recent findings by \cite{gor2017a,gor2017b} and test their proposed fitting model  (i.e. one single component with a spectral break in the low-energy part of the spectrum).
The data extraction and analysis method are described in \S~\ref{sec:method}.
The results obtained from the modelling of time-integrated and time-resolved spectra are reported in \S~\ref{sec:time_integrated} and \S~\ref{sec:time_resolved}, respectively.   
\S~\ref{sec:disc} presents a discussion of the results. The main findings are summarized in \S~\ref{sec:concl}. 

\section{Method}\label{sec:method}

\subsection{Detectors and energy range selection}
The GBM is composed of 12 sodium iodide (NaI, 8\,keV to 1\,MeV) and two bismuth germanate (BGO, 200\,keV to 40\,MeV) scintillation detectors. 
We analysed the data from the two NaI with the highest count rates, namely NaI\,6 and NaI\,9, and from both BGO detectors.
Data were retrieved from the online archive\footref{fn:GBMcatalog}, and analysed using the dedicated software RMFIT~(v.~4.3.2).

We adopted the procedure explained in the Data Analysis Threads and Caveats\footnote{https://fermi.gsfc.nasa.gov/ssc/data/}.
In particular, we selected energy channels in the range 8--900\,keV for NaI detectors, and 0.3--40\,MeV for BGO detectors, and excluded channels in the range 30--40\,keV due to the presence of the Iodine K-edge at 33.17\,keV. 
A free inter-calibration constant factor between the NaI and BGO detectors is included.  
Background spectra are selected in time intervals far from the burst and fitted with a fourth-order polynomial function.
The most updated response matrix files (released on September 6, 2017) have been adopted in our analysis.

GRB~160625B has also been detected  by the LAT on board \fe\ 
\citep[e.g.]{Wang2017,Lu2017}. 
The time-integrated  analysis of the main event has been performed  using GBM data alone and adding, at a later stage, the LAT Low Energy events (LLE data, 30--100\,MeV), performing a simultaneous fit with the RMFIT software.
LAT-LLE data have been extracted from the online \fe\ LLE Catalog\footnote{https://heasarc.gsfc.nasa.gov/W3Browse/fermi/fermille.html}.

\begin{figure}
\centering
\includegraphics[scale=0.28]{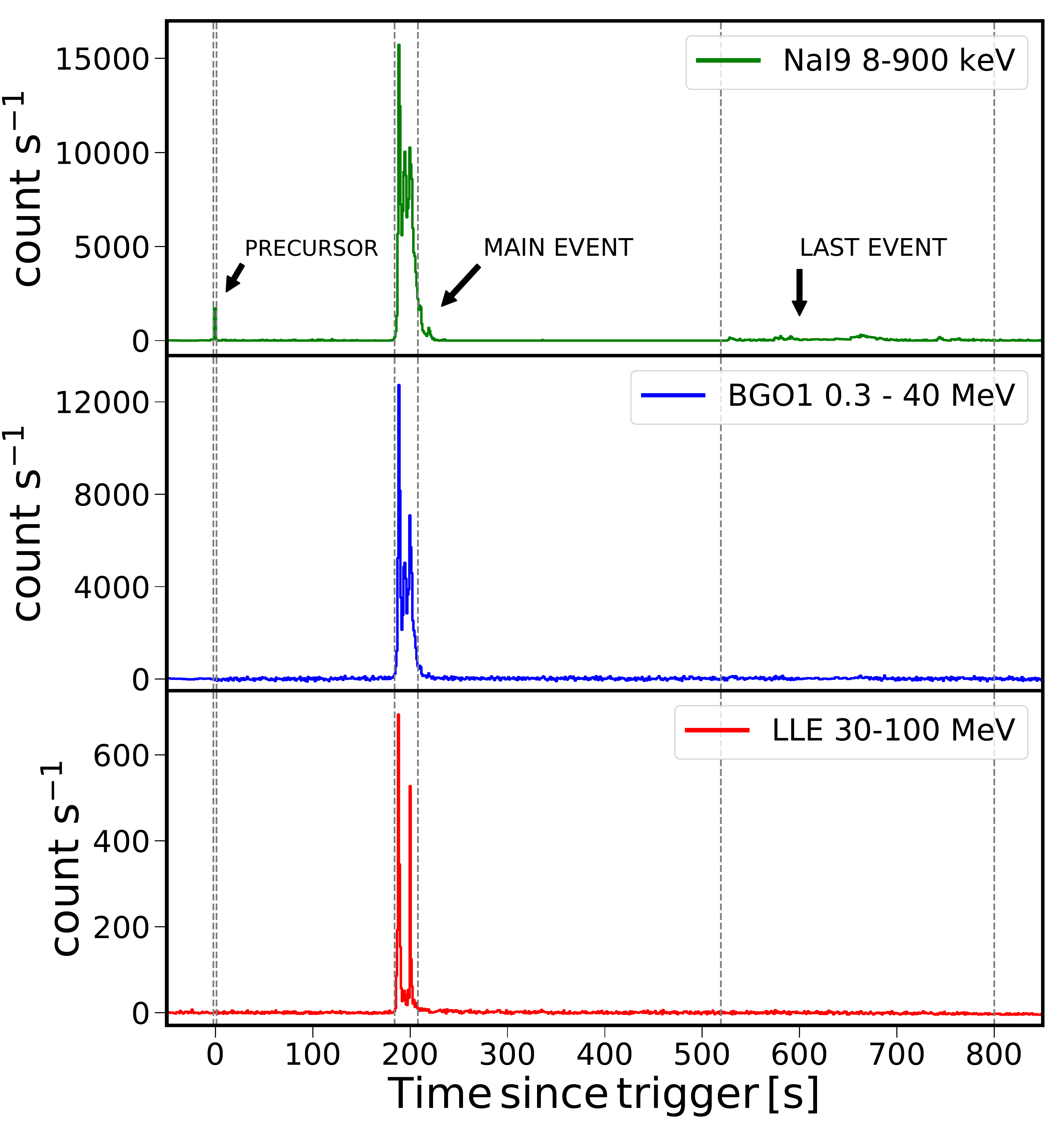}
\caption{Background-subtracted light curves of GRB~160625B detected by NaI\,9  (8\,keV--900\,keV, top), BGO\,1  (300\,keV--40\,MeV, middle), and LAT--LLE (30\,MeV--100\,MeV, bottom).}\label{fig:lc}
\end{figure}

\subsection{Time interval selection}
We analysed the CSPEC data, which provide a time resolution of 1.024\,s and high spectral resolution, comprising 128 logarithmically spaced energy channels. 
Figure~\ref{fig:lc} shows the light curve in three different energy ranges: 8--900\,keV (upper panel, NaI\,9), 0.3--40\,MeV (middle panel, BGO\,1), and 30--100\,MeV (bottom panel, LLE).
Three different emission episodes separated by long quiescent times are visible: a precursor at $T=T_0$, the main event $\sim$\,180\,s later (lasting approximately 30\,s), and a faint, soft, long-lasting ($\sim$\,300\,s) emission starting at $T\sim T_0+$\,500\,s.

We present the results of time-integrated and time-resolved spectral analysis on the main emission episode. Spectral analyses of the precursor and last soft emission episode were performed and are presented in Appendix~\ref{app:prec_and_last}. For all three episodes, the time interval over which the spectrum was accumulated is marked with vertical dashed lines in Fig.~\ref{fig:lc}.

For the main emission episode, the time interval for spectral analysis was selected requiring a signal-to-noise ratio (S/N) higher than 20 in the brightest BGO (BGO\,1).
This criterion results in the selection of the time interval 186.40--207.91\,s. We performed the  analysis of the spectrum integrated over this time interval (time-integrated spectrum) and of the 21 bins (time-resolved spectra), with time integration of 1.024\,s each, distributed within the above time interval.

\subsection{Preliminary analysis: evidence of deviation from standard models}\label{sec:preliminary}

The \fe-GBM GRB Catalog reports the results of the analysis on the spectrum integrated between $T_{\rm start}-T_0=-44$\,s and $T_{\rm stop}-T_0=789$\,s.
According to this analysis,  among a simple power law (PL), a cut-off power law (CPL), a Band model, and a smoothly broken power law (SBPL), the best model  is the last, with $\alpha=-1.021\pm 0.004$,
$E_{\rm peak}=(511 \pm 27)$\,keV,
and $\beta=-2.096 \pm 0.014$.
The reduced chi-square is, however, extremely large, $\chi_{\rm red}^{2} = 4.20$.
This suggests that none of the standard models provides a good fit for this spectrum.

We restrict the analysis to the main emission episode (186.40--207.91\,s).
This choice also gives us the possibility to check if the poor fit is caused by strong spectral evolution from the precursor to the late time soft emission. 
A SBPL function (see equation~\ref{eq:sbpl}) returns $\alpha = -0.722 \pm 0.004$, $E_{\rm peak}= 327.5\pm 2.8$\,keV, $\beta=-2.184 \pm 0.005$, and $\chi_{\rm red}^{2} = 6.51$. The chi-square is again very large, and the fit has not improved. 
The spectrum and SBPL fit are shown in the upper panel of Fig.~\ref{fig:avgsp}, together with the data-to-model residuals (in units of the data standard error).
Residuals are characterized by a systematic trend, with broad excesses peaking around 60\,keV and 600\,keV.

A possible solution to improve the fit, which has been typically considered in the literature in this and in similar cases \citep{Zhang2016,Wang2017,Lu2017}, is to add a BB component. 
We test this possibility and perform a spectral fit with a two-component model,  SBPL+BB, shown in the middle panel of Fig.~\ref{fig:avgsp}.
The chi-square reduces to $\chi_{\rm red}^{2} = 1.97$, and the BB temperature is found at $kT=34.45_{-0.38}^{+0.39}$\,keV, and accounts for the low-energy excess that was evident around 60\,keV in the top panel of Fig.~\ref{fig:avgsp}.
The peak energy of the SBPL component shifts to higher energies (almost a factor of 2, $E_{\rm peak}=576.3_{-5.99}^{+6.15}$\,keV), eliminating the excess that was visible in the one-component SBPL fit at $600$\,keV.
The SBPL photon indices also change considerably after the BB is added, becoming $\alpha=-0.914 \pm 0.005$, $\beta=-2.432 \pm 0.010$.

\cite{gor2017a} suggested that an alternative solution to the presence of a BB is to consider a one-component model with a break in the low-energy part of the spectrum (for a  comparison of these models, see Fig.~\ref{fig:sketch}, green and red curves).
Following the results from \cite{gor2017a,gor2017b}, we test this possibility. 
To this end, we modify the SBPL function to include a low-energy break and an additional PL segment below the break.

\subsection{ Double smoothly broken power law (2SBPL)}\label{sec:sbpl2}
The single-component spectral models traditionally used to fit the GBM spectra (e.g. \citealt{Kaneko2006}) include a PL, a CPL, the Band model, and a SBPL. 
The advantage of the SBPL with respect to the Band model is that it allows  the smoothness of the curvature connecting the two PL segments to be changed. 
The SBPL function used in the GBM Catalog is defined in \cite{Kaneko2006}.
In order to easily extend the definition of the SBPL to more than one break, in this work we start from a different definition: 
\begin{equation}
N^{\rm SBPL}_{\rm E}
= A E_{\rm j}^{\alpha} \Biggl[\Biggl(\frac{E}{E_{\rm j}}\Biggr)^{-\alpha n}+\Biggl(\frac{E}{E_{\rm j}}\Biggr)^{-\beta n}\Biggr]^{-\frac{1}{n}}~,
\label{eq:sbpl}
\end{equation}
where
\begin{equation}
E_{\rm j} = E_{\rm peak} \cdot \Biggl(- \frac{\alpha + 2}{\beta + 2}\Biggr) ^{\frac{1}{(\beta - \alpha)n}} ~.
\end{equation}
In Eq. (1) $N_{\rm E}$ is the photon spectrum (i.e. number of photons per unit area, per unit time, and per unit energy). 
The free parameters are the amplitude $A$, the low-energy spectral index $\alpha$, the peak energy of the $E^2N_{\rm E}$ spectrum $E_{\rm peak}$, the high-energy spectral index $\beta$, and the smoothness parameter $n$ (higher values of $n$ correspond to sharper curvatures).

In the GBM Catalog the smoothness parameter is called $\Lambda$ and is kept fixed to $\Lambda=0.3$ for all GRBs (see \citealt{Kaneko2006} for an explanation).
In order to perform a fit that can be compared to the one reported in the GBM Catalog, our smoothness parameter $n$, which has a different definition, has been fixed to the value $n=2.69$, corresponding to $\Lambda=0.3$.

We extend the SBPL mathematical function to include a second break energy and a third power-law segment:
\begin{multline}
N^{\rm 2SBPL}_{\rm E} = A \, E_{\rm break}^{\alpha_1} \, \Biggl[ \, \Biggl[ \Biggl(\frac{E}{E_{\rm break}}\Biggr)^{- \alpha_1 n_1}+\Biggl(\frac{E}{E_{\rm break}}\Biggr)^{- \alpha_2 n_1}\Biggr]^{\frac{n_2}{n_1}}+\\
+\Biggl(\frac{E}{E_{\rm j}}\Biggr)^{- \beta \, n_2} \cdot \Biggl[\Biggl(\frac{E_{\rm j}}{E_{\rm break}}\Biggr)^{- \alpha_{1} n_{1}}+\Biggl(\frac{E_{\rm j}}{E_{\rm break}} \Biggr)^{- \alpha_2 n_1} \Biggr]^{\frac{n_2}{n_1}} \Biggr]^{-\frac{1}{n_2}}~,
\label{eq:2sbpl}
\end{multline}
where
\begin{equation}
E_{\rm j} = E_{\rm peak} \cdot \Biggl(- \frac{\alpha_2 + 2}{\beta + 2}\Biggr) ^{\frac{1}{(\beta - \alpha_2) \, n_2}}~.
\end{equation}

\begin{figure}[t]
\centering
\includegraphics[scale=0.45]{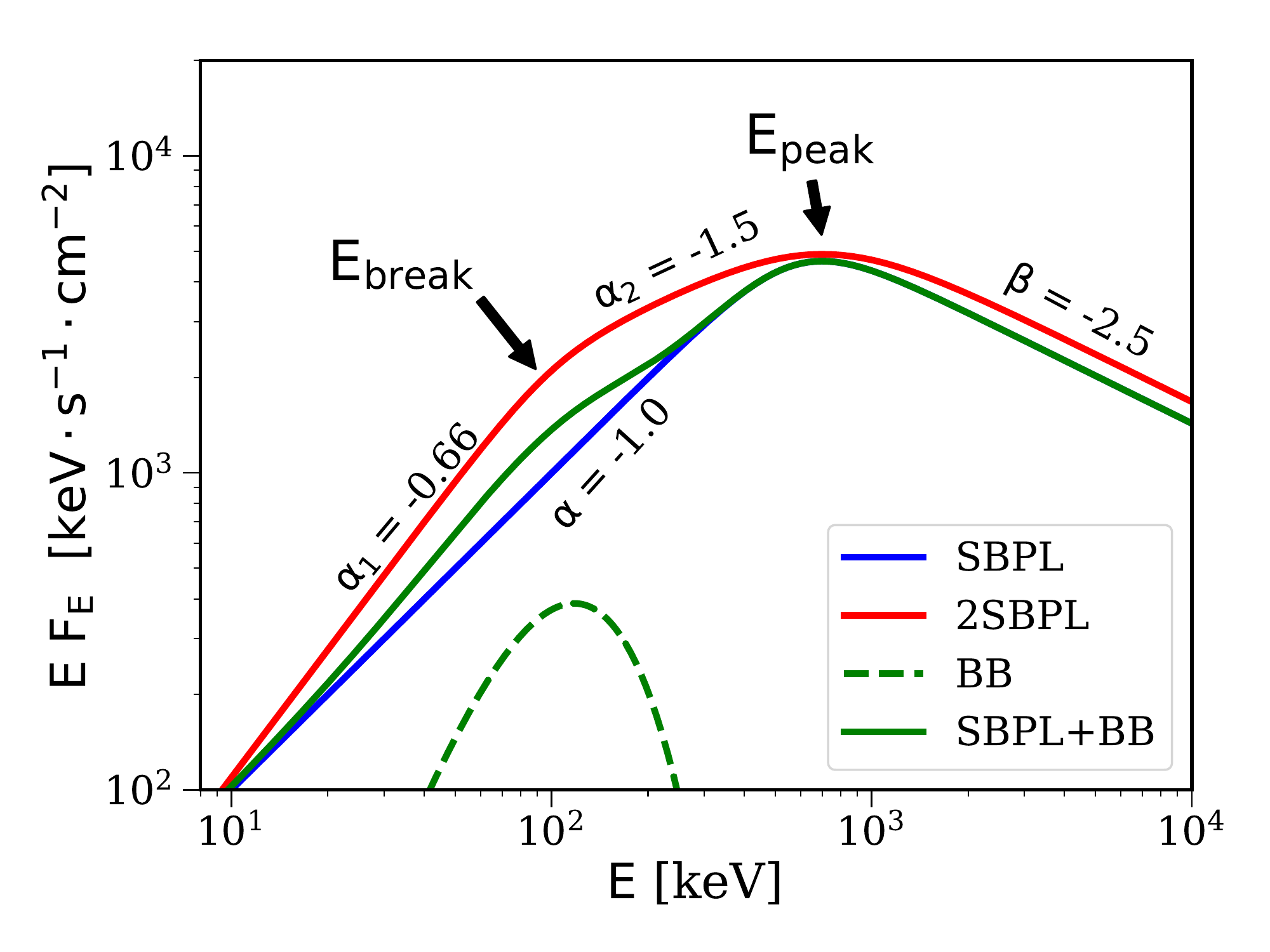}
\caption{Comparison between the SBPL model (blue curve), SBPL+BB (green solid curve), and 2SBPL (red curve). Normalizations are arbitrary.}
\label{fig:sketch}
\end{figure}

The free parameters are the amplitude $A$, the photon index $\alpha_1$ below the break energy, the break energy $E_{\rm break}$, the photon index $\alpha_2$ between the break and the peak energies, the peak energy $E_{\rm peak}$, the high-energy photon index $\beta$, and the smoothness parameters $n_1$ (for the break) and $n_2$ (for the peak).

As before, we fix the curvature around the peak energy to the value $n_2=2.69$.
After performing time-resolved spectral fitting by leaving $n_1$ free to vary, we realized that the model parameters of the fit are not always constrained, and so we also decided to fix  the value of $n_1$.
We fix $n_1$ to the mean value of the distribution inferred when it is left free to vary: $n_1=5.38$. This corresponds to a sharper curvature than the curvature around the peak energy.

The 2SBPL model (Eq.~\ref{eq:2sbpl}) is nested into the SBPL (Eq.~\ref{eq:sbpl}).
The fits obtained from the two models can then be compared through an $F$-test \citep{Protassov2002}. 
We implemented the SBPL\footnote{For the reasons discussed in e.g. \cite{Cabrera2007} and \cite{Calderone2015}, the version of the SBPL we implemented in the library of RMFIT uses $E_{\rm peak}$ as free parameter rather than $E_{\rm j}$, so that the result of the fit gives directly $E_{\rm peak}$.} and the 2SBPL in RMFIT.

These models are shown (assuming typical parameters for the photon indices) in Fig.~\ref{fig:sketch} (SBPL in blue and 2SBPL in red).
For comparison, we also show a SBPL+BB (green line).
As is evident, the overall effect of adding a (non-dominant) BB is similar to the effect of considering a softer SBPL (i.e. more consistent with synchrotron, $\alpha_2=-1.5$) and adding a break at low energies.  The final functions have a similar shape (red and green solid lines in Fig.~\ref{fig:sketch}).

\section{Time-integrated analysis}\label{sec:time_integrated}
\begin{figure}[t]
\centering
\includegraphics[scale=0.156,trim=3cm 0 3.2cm 0]{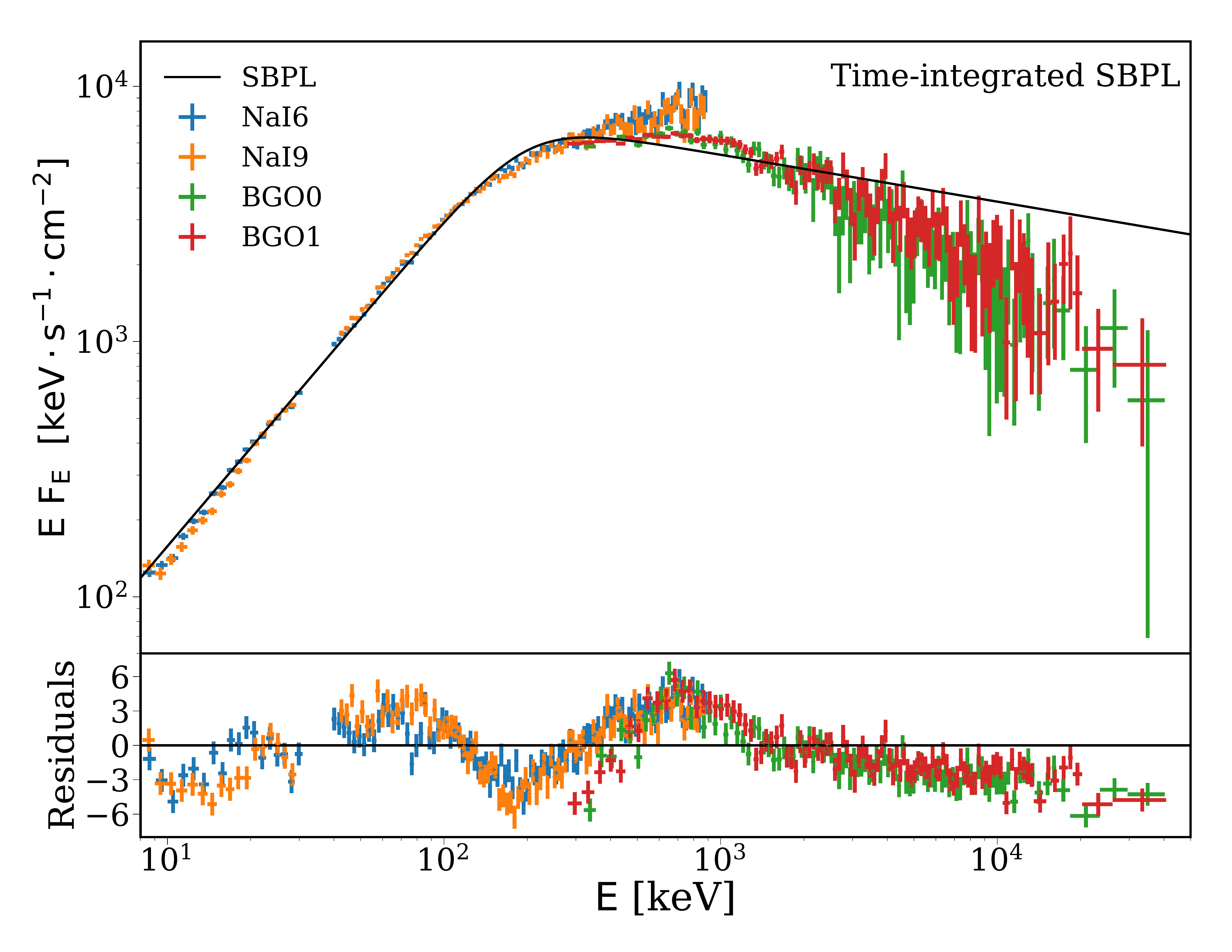}
\includegraphics[scale=0.156,trim=3cm 0 3.2cm 0]{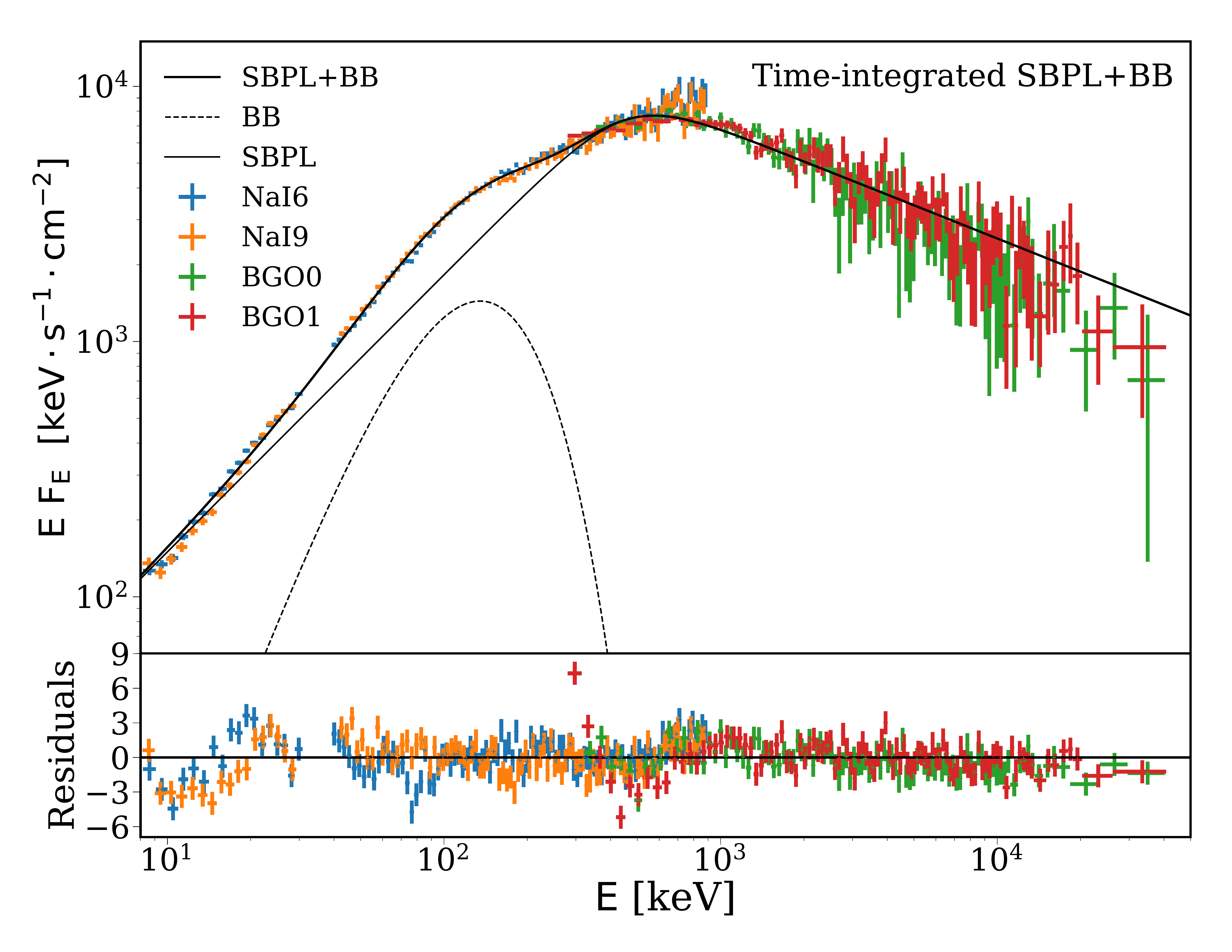}
\includegraphics[scale=0.156,trim=3cm 0 3.2cm 0]{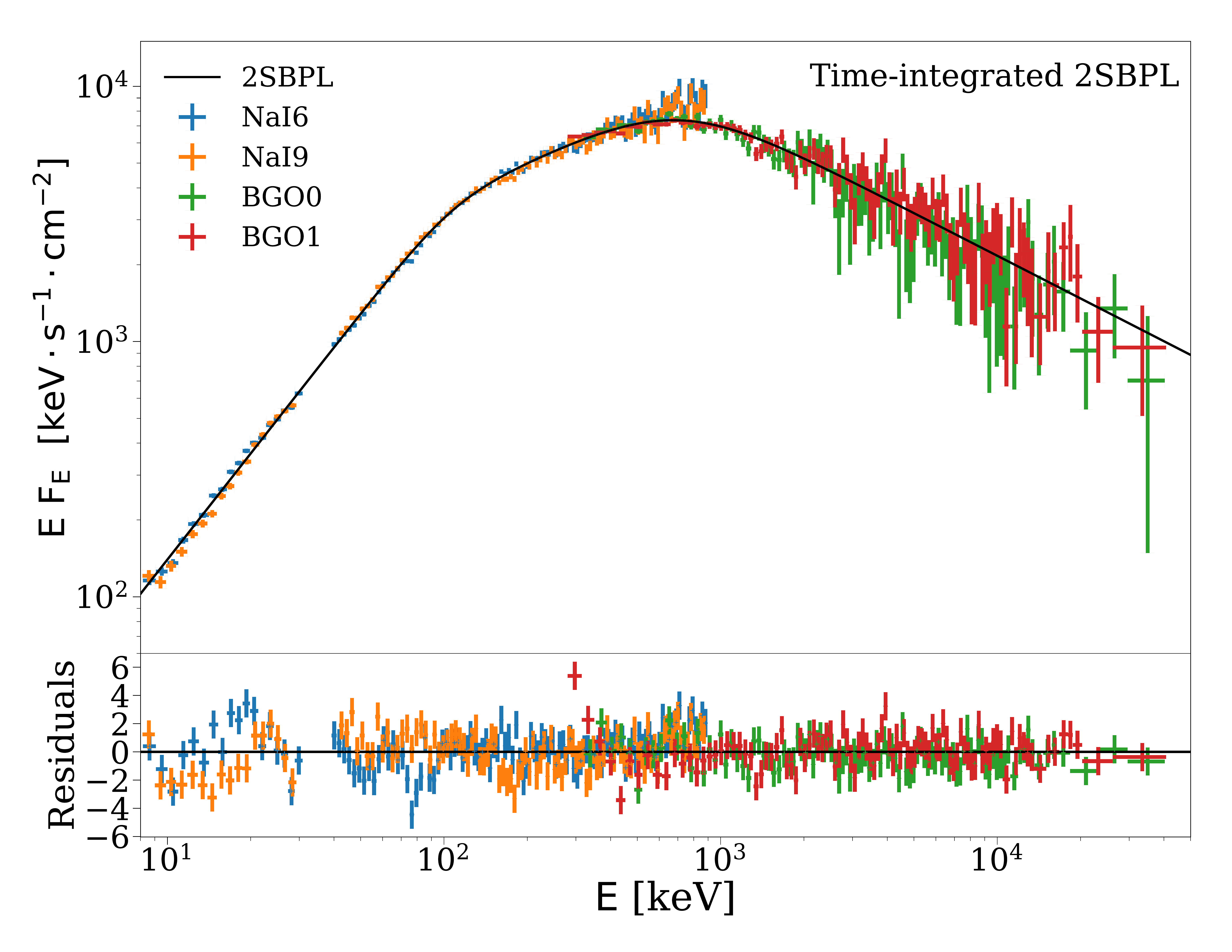}
\caption{Time-integrated spectrum of the main event (186.40--207.91\,s). Three different models are tested:  SBPL, SBPL+BB, and 2SBPL (from top to bottom). Different colours refer to different instruments, as explained in the legend. 
 In each panel, the bottom stripe 
  shows the model residuals.}
\label{fig:avgsp}
\end{figure}

We fit the 2SBPL function, defined in equation~\ref{eq:sbpl}, to the time-integrated spectrum of the main emission episode (time interval 186.40--207.91\,s).
The result is shown in the bottom panel of Fig.~\ref{fig:avgsp}. 
The chi-square is $\chi_{\rm red}^2=701.9/462=1.52$, corresponding to an improvement at more than 8$\sigma$  compared to the SBPL fit.

A spectral break is found at $E_{\rm break} = (107.8\pm1.9)$\,keV.
The peak energy increases (compared to previous tested models) to $E_{\rm peak} = 673.5 \pm 10.8$\,keV.
The photon indices below and above \eb\ have best fit values $\alpha_1=-0.62\pm0.01$ and $\alpha_2=-1.50\pm0.01$, respectively. 
These values are very close to those expected from synchrotron emission from a cooled population of electrons. 

We recall that the same spectrum, when modelled with a SBPL+BB (section~\ref{sec:preliminary}) gives $\chi_{\rm red}^2=909.7/462=1.97$.
Since the SBPL+BB and 2SBPL are not nested models, but  have the same number of degrees of freedom, they can be compared in terms of $\chi^2$ and associated probability. 
This comparison  favours the 2SBPL model.
However, we note that both fits have a large reduced chi-square. 
The main contribution comes from the inconsistency between the two NaI, especially at low energies (i.e. in some energy ranges, one is systematically above/below the other).

\begin{figure}
\centering
\includegraphics[scale=0.156]{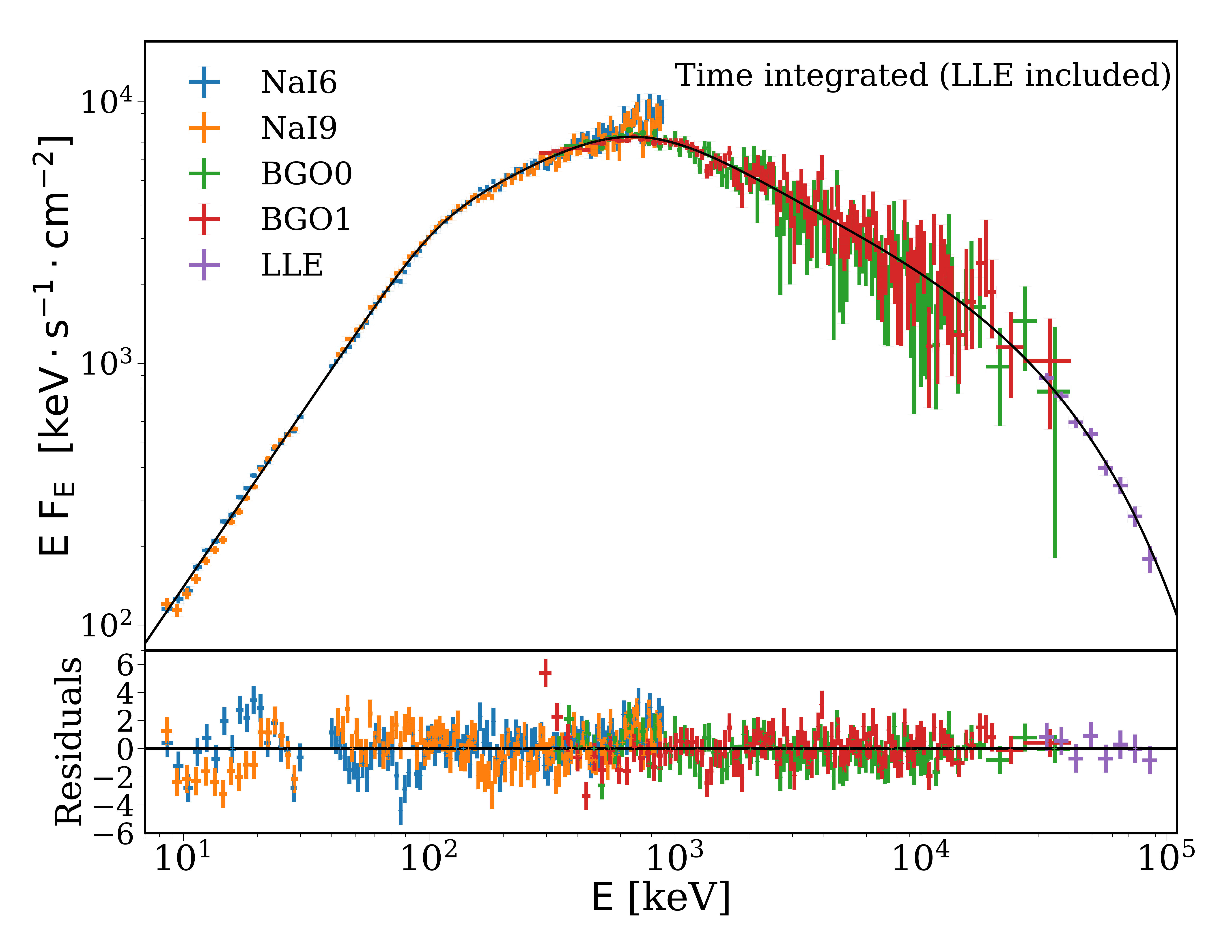}
\caption{Time-integrated spectrum (186.40--207.91\,s) from 8\,keV to 100\,MeV, including LAT-LLE data. The model (black solid line) is a 2SBPL with a high-energy exponential cut-off.}
\label{fig:lat}
\end{figure}

Since in the time interval we are considering for the time-integrated analysis, LAT observations are also available, it is worth investigating their consistency with the GBM data. 
We find the LLE data do not lie on the extrapolation of the BGO data: they instead reveal the presence of a softening at high energies.
In order to model this softening, we modify the 2SBPL by adding an exponential cut-off at high energy.
The fit shown in Fig.~\ref{fig:lat} with the solid black line.
The LLE data are shown with purple symbols. 
The best fit value of the cut-off energy (defined as the energy at which the flux is suppressed by a factor $\sim1/e$ as compared to the simple PL extrapolation) 
 is $E_{\rm cut}=50.3_{-13.2}^{+7.4}$\,MeV, and the reduced chi-square is $\chi_{\rm red}^2=1.51$.
All the other spectral parameters (photon indices, low-energy break and peak energies) are consistent with those obtained when LLE data are not included: $\alpha_1 = -0.62 \pm 0.01$, $E_{\rm break}= 107.3_{-1.6}^{+1.9}$, $\alpha_2= -1.49 \pm 0.02$, $E_{\rm peak} = 668.7_{-9.2}^{+14.4}$, and $\beta=-2.54_{-0.02}^{+0.03}$.
If interpreted as being  caused by photon-photon annihilation, the cut-off at $\sim50$\,MeV corresponds to a Lorentz factor $\sim200-250$, for a variability timescale $\sim 1-0.1$\,s \citep{lithwick01}.
Similar cut-off energies have been identified, from simultaneous modelling of GBM and LLE data, by \cite{vianello17}.

We also  tested  a model where the high-energy softening is modelled by an additional power law instead of an exponential cut-off, i.e. a model with
four power-law segments connected smoothly by three breaks. This fit has a reduced chi-square $\chi_{\rm red}^2=1.51$, equal to that  obtained when an exponential cut-off is used. 
The highest energy break (i.e. above the peak energy) is found at $27.3_{-6.7}^{+4.8}$\,MeV. The power-law segment above the high-energy break has photon index $-3.5\pm0.2$.
The values of the other parameters are similar to those found when the softening is modelled using an exponential cut-off. If we interpret the break at 27.3\,MeV as being due to photon-photon annihilation, we estimate a $\Gamma\sim 150-200$, and the MeV-GeV power-law segment can be explained by emission and absorption taking place in the same region.

\section{Time-resolved spectral analysis}\label{sec:time_resolved}

In order to check whether the low-energy break identified in the time-integrated spectrum is also present  in the time-resolved spectra and study its evolution with time, we divided the time interval 
186.40--207.91\,s into 21 time bins, with 1.024\,s integration each. 
As for the time-integrated analysis, we tested the three following models: SBPL, 2SBPL, and SBPL+BB (see Fig.~\ref{fig:sketch}).
The results obtained from the different models (best fit parameters, photon and energy flux, chi-square and degrees of freedom, fit probability, and $F$-test) are listed in Table~\ref{tab:sbpl} (SBPL), Table~\ref{tab:2sbpl} (2SBPL), and Table~\ref{tab:sbplBB} (SBPL+BB).

First, we compared through an $F$-test the SBPL and 2SBPL models for all 21 time-resolved spectra.
In 19 spectra (i.e. all spectra except the last two, where the flux is small), the 2SBPL improves the SBPL fit at more than 3$\sigma$, which we take as the threshold value for the definition of the best fit model.
More specifically, in all these 19 spectra the fit improves at more than 4.8$\sigma$ (more than 8$\sigma$ in 13 cases).

Before presenting the temporal evolution of the 2SBPL best fit parameters, we comment on the SBPL+BB fits.
 The SBPL+BB model also leads to a significant improvement of the fit over the SBPL.
A comparison between SBPL and SBPL+BB in terms of $F$-test, however, cannot be performed.

\begin{figure}[t]
\centering
\includegraphics[scale=0.48,trim= 0 -0.5cm 0.6cm -0.3cm ]{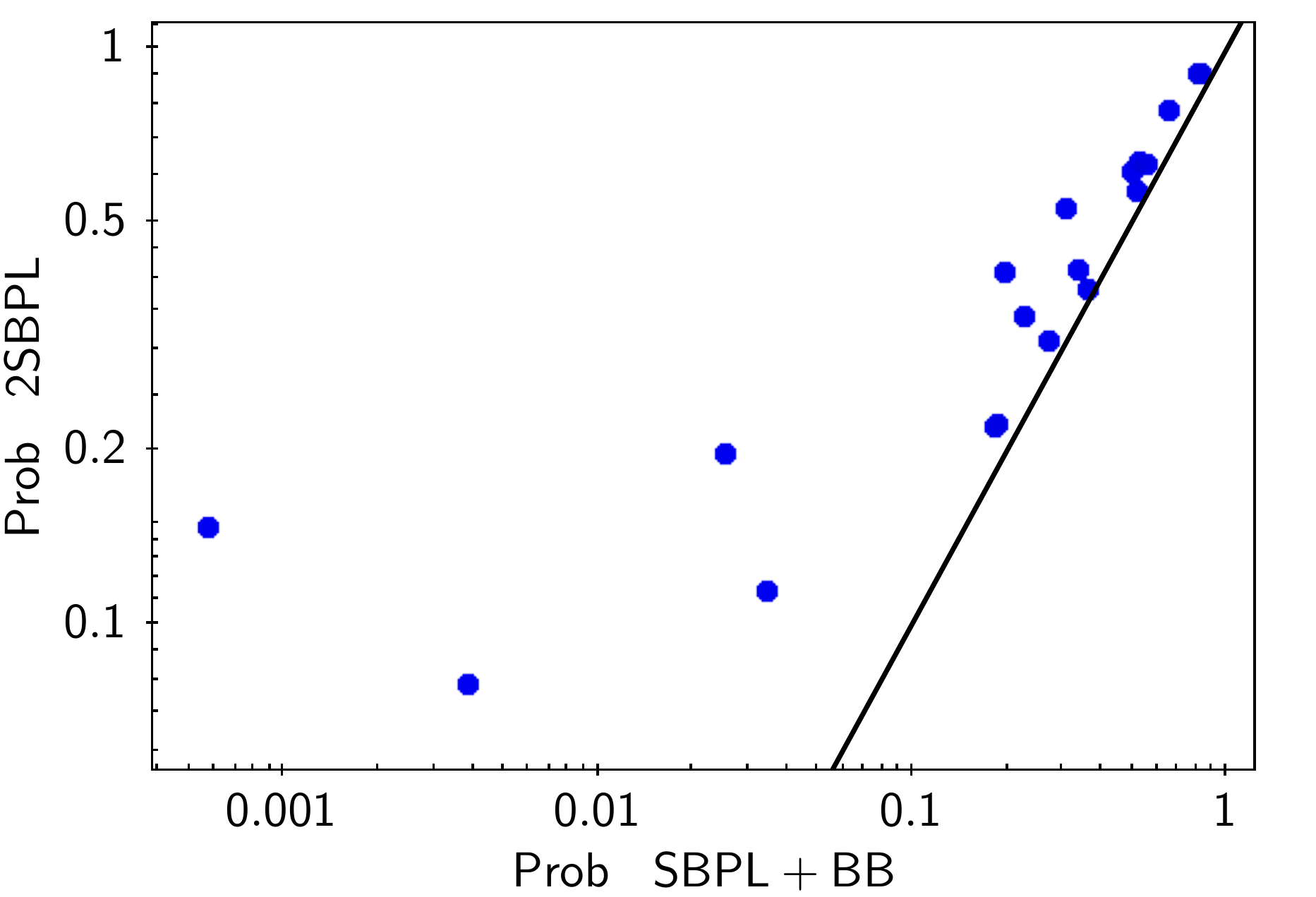}
\caption{Fit probability for time-resolved spectra: comparison between fits performed with a 2SBPL model ($y$-axis) and with a SBPL+BB model ($x$-axis). The two models have the same number of degrees of freedom. The equality line is shown as a solid black line.}
\label{fig:prob}
\end{figure}

A comparison between 2SBPL and SBPL+BB can  instead be performed in terms of probability of their $\chi^2$ since they are not nested models, but  have the same number of parameters and degrees of freedom. 
The chi-square probability is reported in Table~\ref{tab:2sbpl} and Table~\ref{tab:sbplBB} for each time-resolved spectrum, for the 2SBPL and SBPL+BB models, respectively. 
The two probabilities are compared in Fig.~\ref{fig:prob}, where the equality line is shown as a solid black line.
The probabilities of the 2SBPL are systematically larger than those resulting from an SBPL+BB fit. 
The fits with the 2SBPL and with the SBPL+BB shift the spectral peak at slightly higher values than  the SBPL: this result has been found in all the time-resolved spectra (see Appendix~\ref{app:tables}). 
Notably, the break energy of the 2SBPL coincides with the peak of the BB component.

\begin{figure}[ht]
\centering
\includegraphics[scale=0.16,trim=0 1.5cm 0 0]{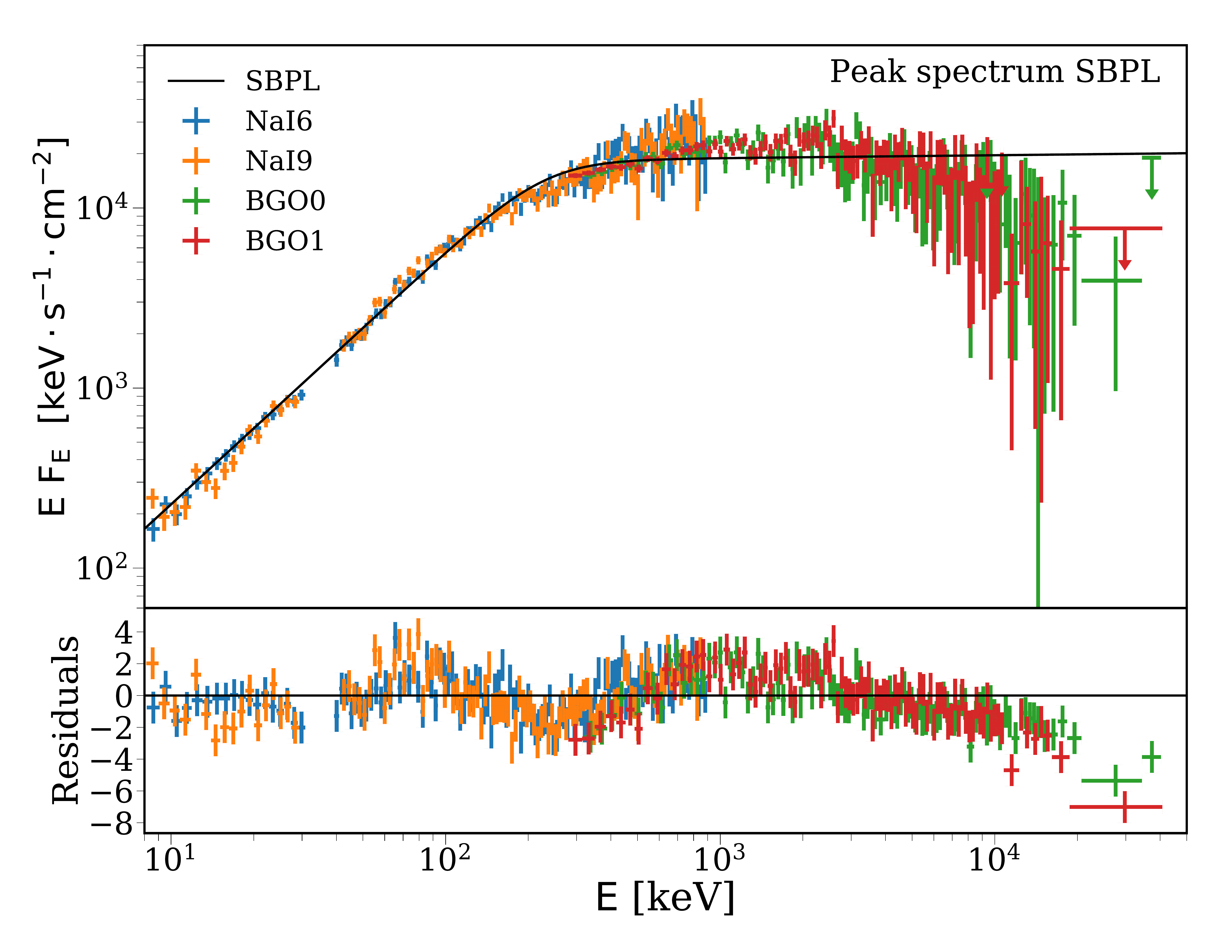}
\includegraphics[scale=0.16,trim=0 1.5cm 0 0]{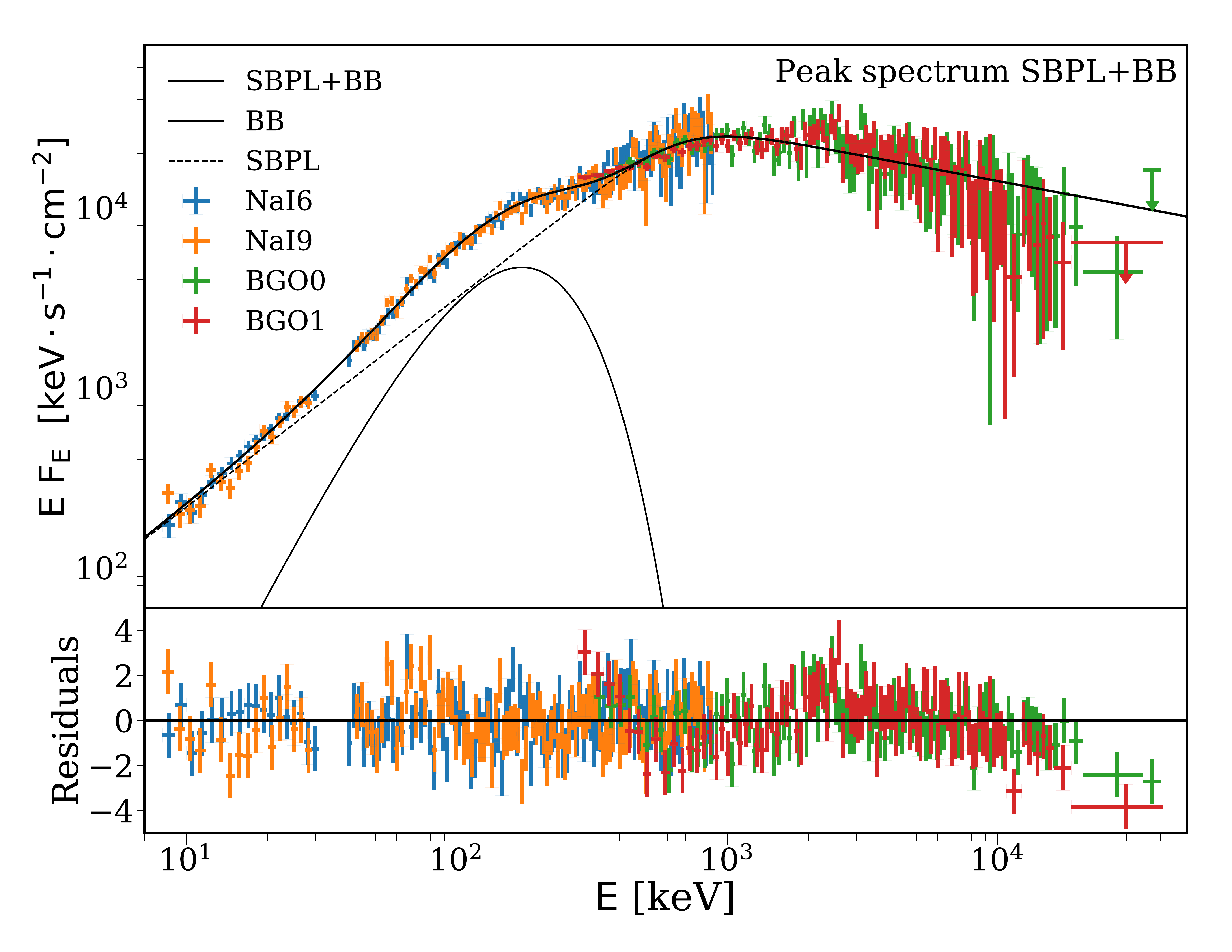}
\includegraphics[scale=0.16,trim=0 3.cm 0 0]{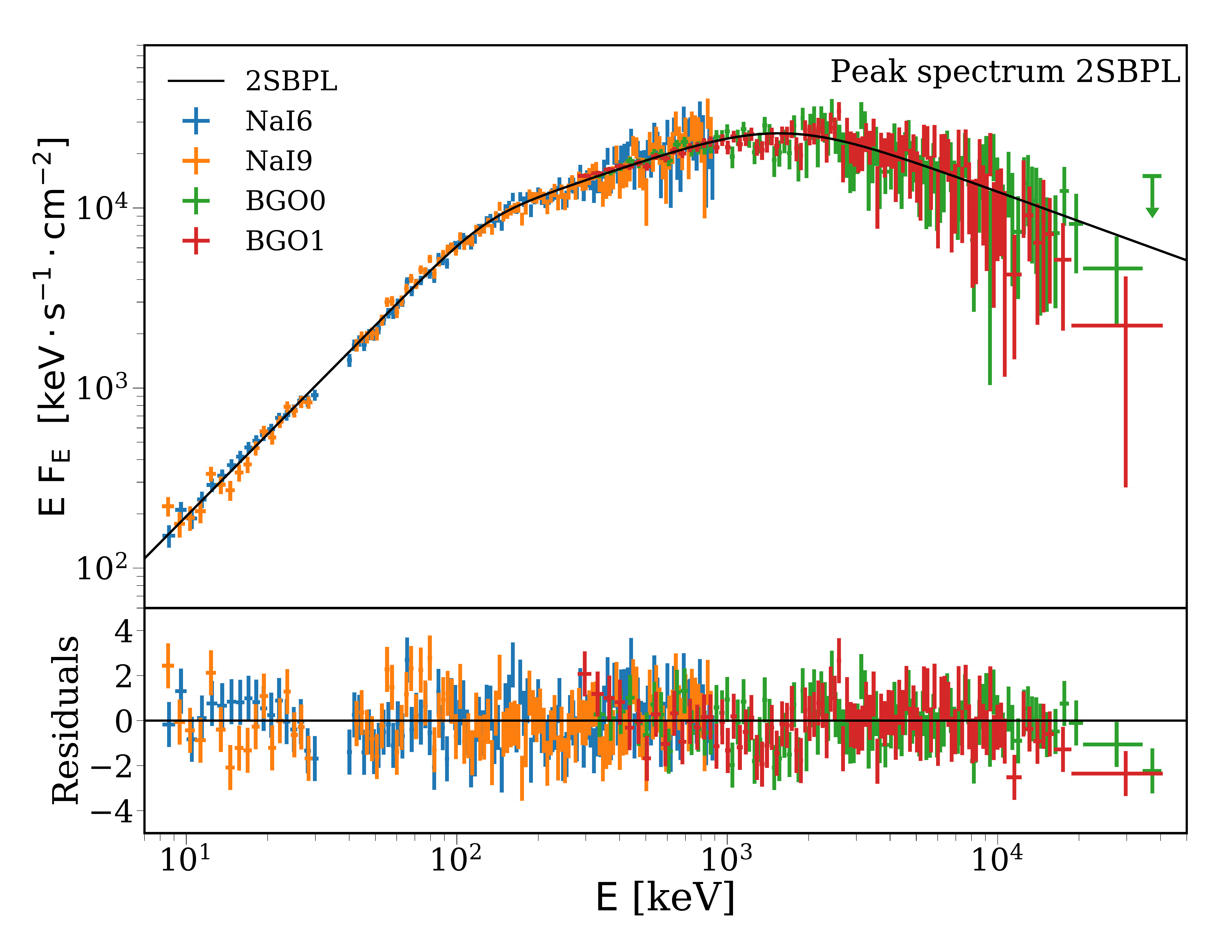}
\caption{Time-resolved spectrum accumulated in the time interval 188.45--189.47\,s  (peak of the light curve). Different spectral models are tested: a SBPL (upper panel), SBPL+BB (middle), and 2SBPL (lower panel). }
\label{fig:sp_main}
\end{figure}

As an example,  in Fig.~\ref{fig:sp_main} we show the results obtained from the spectrum at the peak of the light curve (time bin 188.45\,s--189.47\,s).
Similarly to the time-integrated spectrum, a SBPL model (top panel) has large residuals, displaying a characteristic trend.
The situation largely improves both when a BB component is added (middle panel) and when the SBPL is modified to have one additional PL (2SBPL, bottom panel).

In the peak spectrum the statistical comparison based on the chi-square firmly favours 
 a 2SBPL model over the SBPL+BB: $\chi^2_{\rm 2SBPL}=1.07$ ($P_{\rm 2SBPL}=0.15$) and $\chi^2_{\rm SBPL+BB}=1.23$ ($P_{\rm 2SBPL}=6\times10^{-4}$).
In particular, the SBPL+BB function seems to underestimate the energy of the spectral peak ($E_{\rm peak}^{\rm SBPL+BB}\sim1$\,MeV), which is instead better modelled by the 2SBPL function ($E_{\rm peak}^{\rm 2SBPL}\sim1.5$\,MeV).

We give a tentative explanation of  why in this spectrum the data favours the 2SBPL model over the SBPL+BB. The peak spectrum is the spectrum that simultaneously maximizes two conditions: \ep\ and \eb\ should be well separated (with a ratio $\gtrsim10$, see bottom panel in Fig.~\ref{fig:timevo}) and the photon statistics should be large (this bin corresponds to the peak flux of the light curve).
In the SBPL+BB model, the BB is used to model the break energy and the peak of the non-thermal component is used to model the peak of the spectrum. 
When these two features are far from each other, the SBPL+BB model predicts a dip between the BB peak and the SBPL peak.
If such a dip is not present in the data, the SBPL+BB model is then forced (in order to minimize the chi-square) to lower the value of the SBPL peak energy in order to keep large the flux between 500\,keV and 1\,MeV. However, this will result in a slightly worse modelling of the data around the spectral peak.

We then suggest that bright spectra with high ratios between \ep\ and the low-energy hardening are the best cases to distinguish between the two competing models.

\subsection{Spectral evolution}
\begin{figure}[t]
\hskip-0.7cm
\centering
\includegraphics[scale=0.26]{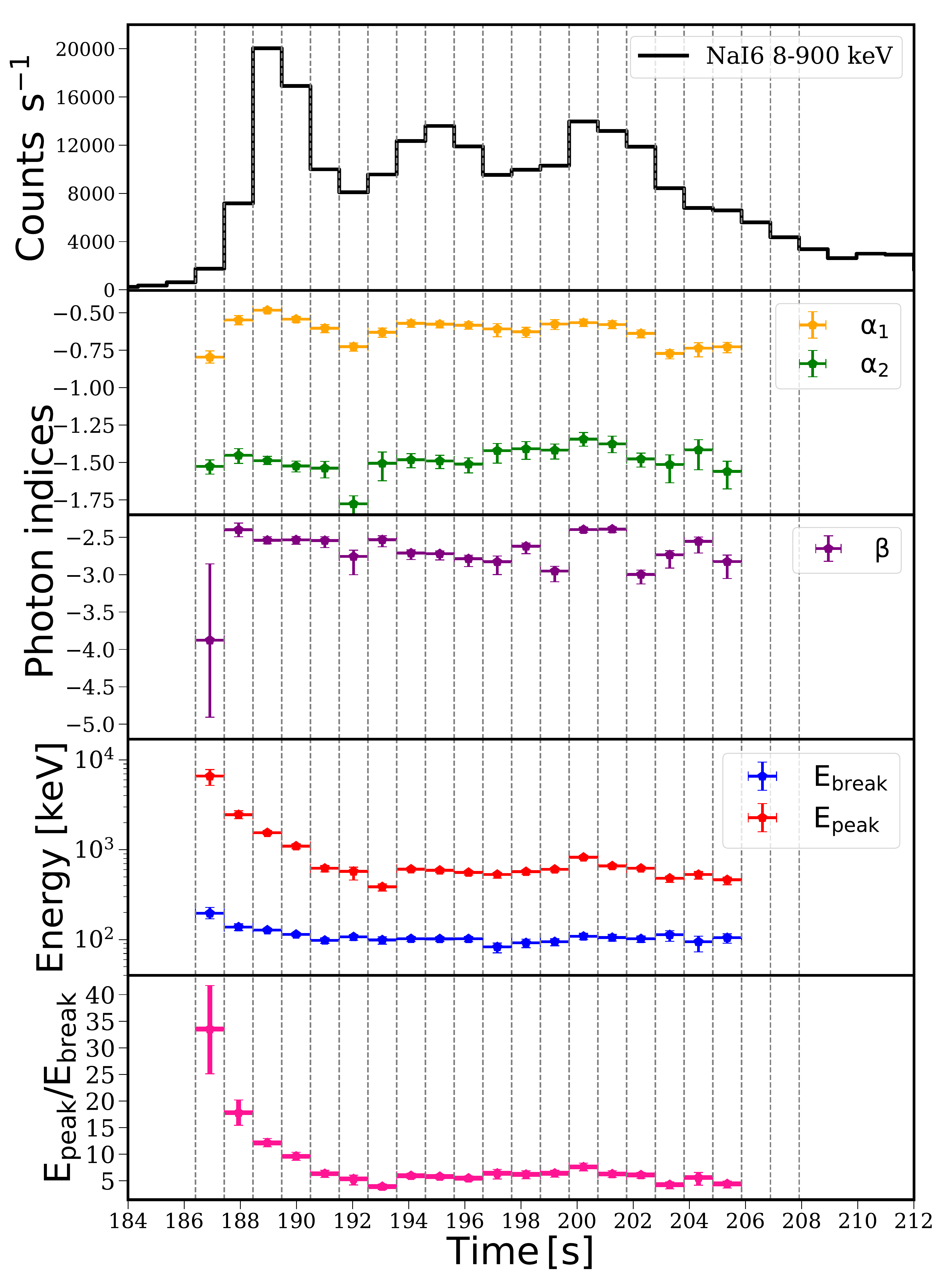}
\caption{Time evolution of the spectral parameters of the 2SBPL model (Table~\ref{tab:sbpl}) for time-resolved spectra where the 2SBPL fit improves at more than 3$\sigma$ the SBPL fit (all bins but the last two). 
From top to bottom: Count rate light curve (with 1.024\,s time resolution), photon indices below and above the break (yellow and green symbols, respectively), spectral index above the peak energy (purple symbols), peak and break energy (red and blue symbols, respectively), and ratio between peak and break energy (pink symbols). 
For an explanation of the notation used for the 2SBPL parameters, see Fig.~\ref{fig:sketch} (red line). }
\label{fig:timevo}
\end{figure}

The temporal evolution of the spectral parameters inferred from the 2SBPL fits are reported in Fig.~\ref{fig:timevo}.
The upper panel shows the light curve of the main emission episode with a 1.024\,s temporal resolution.
The vertical dashed lines denote the time bins selected for time-resolved spectral analysis.
In the second and third panel, the evolution of the photon indices are displayed.
The fourth panel shows the temporal evolutions of \ep\ (red symbols) and \eb\ (blue symbols).
Their ratio is given in the bottom panel.

\ep\ exhibits a strong evolution (a softening) in the first 5 seconds, after which it settles to a nearly constant value (\ep$\sim 500-600$\,keV).
\eb\ displays a similar evolution, but the initial softening is much less pronounced.
After the first few seconds, \eb\ also displays a nearly constant behaviour (\eb$\sim 100$\,keV).
The ratio $E_{\rm peak}/E_{\rm break}$ varies from $\sim 35$ at the very beginning to $\sim$5 at later times.

\begin{figure}
\includegraphics[scale=0.38]{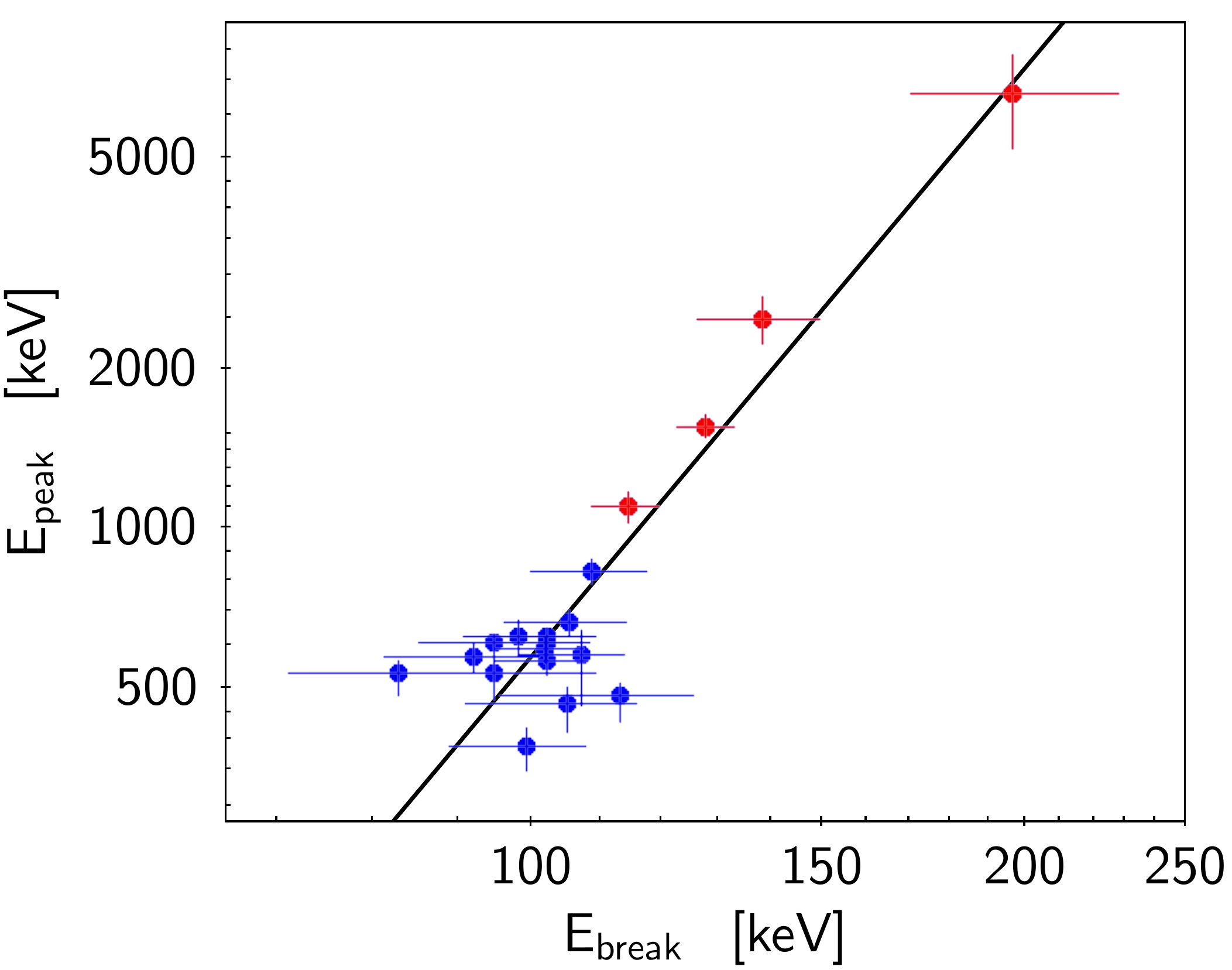}
\caption{Correlation between the peak energy \ep\ and the break energy \eb\ of the 2SBPL model in the time-resolved analysis. 
The values from the first four time bins are indicated by red symbols, while  later times are by blue symbols.
Error bars show uncertainties at 1$\sigma$.}
\label{fig:corr}
\end{figure}

We investigate the presence of a correlation between \ep\ and \eb;
\ep\ versus \eb\ is shown in Fig.~\ref{fig:corr}.
The Spearman correlation coefficient is $\rho=0.61$, with a chance probability $P=0.009$.
Assuming a power-law model, we find 
\begin{equation}
\frac{E_{\rm peak}}{700 \,\rm keV}=(0.81\pm0.06) \left(\frac{E_{\rm break}}{100\,\rm keV}\right)^{3.69\pm0.26}. 
\end{equation}

The power-law fit is shown in Fig.~\ref{fig:corr} by a solid black line.

\begin{figure*}
\centering
\includegraphics[scale=0.5]{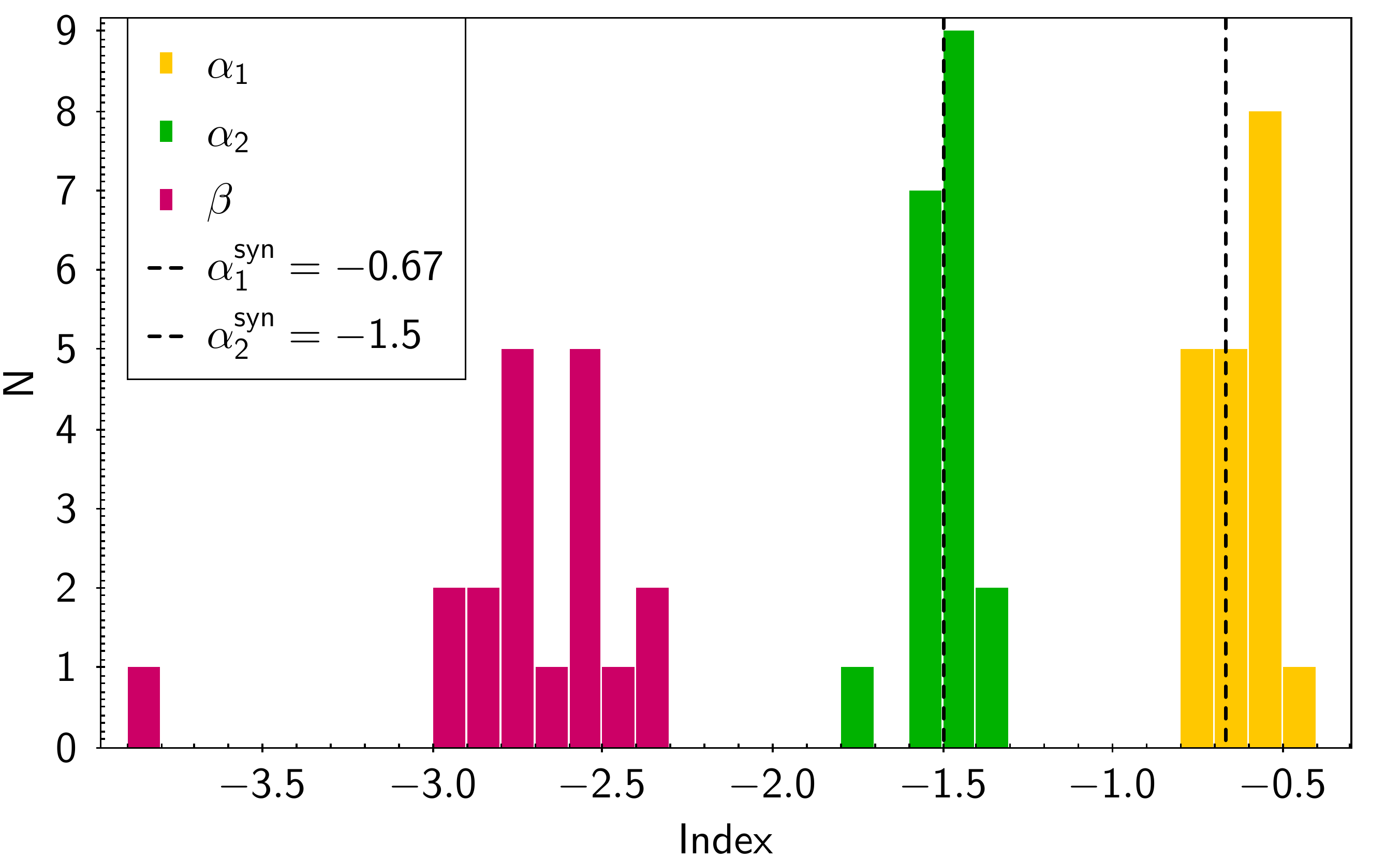}
\caption{Distributions of the spectral indices of the three power-law segments of the 2SBPL model (values in Table~\ref{tab:2sbpl}). 
The vertical dashed lines are the expected values for synchrotron emission in the fast cooling regime.}
\label{fig:indici}
\end{figure*}

Figure~\ref{fig:indici} shows the distributions of the spectral indices of the 2SBPL model fits. 
If modelled with Gaussian functions, the mean values are $\langle\alpha_1\rangle= -0.63$ ($\sigma=0.08$) and $\langle\alpha_2\rangle= -1.48$ ($\sigma=0.09$). 
These values are remarkably consistent with standard synchrotron fast cooling emission, predicting $\alpha_1^{\rm syn}= -2/3$ and $\alpha_2^{\rm syn}= -3/2$.\\


For completeness we have also  analysed  the spectrum of the precursor and of the last dim/long emission episode.
We find that for both episodes the best fit model is a CPL.
Data analysis and results are given in Appendix~\ref{app:prec_and_last}.

\section{Discussion}\label{sec:disc}

In the majority of the spectra of the main emission episode of GRB160625B, an SBPL+BB model returns a similarly good fit to that of the 2SBPL model.
We argued that a possible way to distinguish  between the two models is to consider spectra with a large flux and a high ratio between the peak energy and the low-energy feature.  In this case, if no dip is present in the data between the low-energy feature and the spectral peak, the 2SBPL model will fit this intermediate region with a PL, and will be able to satisfactorily model the spectral peak. The SBPL+BB model will instead be forced to place the peak of the SBPL component at lower energies (compared to the location of the spectral peak), resulting in a larger chi-square.

Another difference between the two models is in their connection to physically motivated models.
The 2SBPL returns photon indices that are remarkably close to those expected in synchrotron spectra. Moreover, this model does not require to invoke the interplay between two different components: one single component explains the data from keV to MeV energies.
Instead, in the competing SBPL+BB model, the SBPL component does not have a straightforward interpretation, since its low-energy photon index has a distribution peaked at $\langle\alpha\rangle= -0.88$ ($\sigma = 0.21$). 
We also tested how the two models compare when their low-energy power-law slope is fixed to the value expected for synchrotron emission, i.e. $-2/3$. While most (90\%) of the fits with the 2SBPL$_{-2/3}$ have a $\chi^2$ probability $>10^{-2}$, and the best fit parameters are similar to those obtained when the low-energy spectral index is left free to vary, in the case of the SBPL+BB$_{-2/3}$ only  45\% of the time-resolved spectra  have a probability $>10^{-2}$.
Therefore, for our discussion, we assume that the spectrum is produced by optically thin synchrotron radiation, and we derive implications on the properties of the magnetic field in a simple standard scenario.

Within this scenario, we identify \eb\ with the cooling frequency $\nu_{\rm c}$ and \ep\ with the 
characteristic synchrotron frequency $\nu_{\rm m}$.
We show in Fig.~\ref{fig:timevo} that \eb\ and \ep\ are fairly close to each other. 
Figure~\ref{fig:timevo} (bottom panel) shows the time evolution of the ratio $E_{\rm break} / E_{\rm peak} \sim 5$. 
The relatively small ratio between \ep\ and \eb\ corresponds to what is generally referred to as a marginally fast cooling regime, i.e. a situation in which $\nu_{\rm cool} \lesssim \nu_{\rm min}$ \citep{Daigne11,Beniamini2013}. This regime, which is still efficient in terms of radiated energy \citep{Daigne11}, requires a relatively long cooling timescale.

\cite{Ghisellini2000} argued that the radiative cooling timescale of electrons is much shorter than the dynamical timescale.
In this case, between the cooling frequency $\nu_{\rm c}$ and the characteristic synchrotron frequency $\nu_{\rm m}$, the flux density scales as $F_\nu \propto \nu^{-1/2}$. 
Considering a population of electrons of energy $\gamma m_{\rm e}c^2$, within a shell moving with bulk Lorentz factor $\Gamma$ and with comoving magnetic field $B^\prime$, the radiative cooling timescale (in the observer frame) for synchrotron emission is
\begin{equation}
t_{\rm cool}^{\rm obs} = {\gamma\over \dot{\gamma}} \, {(1+z)\over \Gamma}  = 
\frac{6 \pi m_{\rm e} c }{\sigma_{\rm T} B^{\prime 2} \gamma (1+ U_{\rm rad}/U_{\rm B})}\,
{(1+z)\over \Gamma}~,
\label{eq1}
\end{equation}
where $U_{\rm rad}$ and $U_{\rm B}$ are the radiation and magnetic energy densities, respectively. 
Our time-resolved spectral analysis was performed considering spectra with an integration time of 1.024\,s. 
Therefore, we can derive a limit on $B^\prime$ by requiring $t_{\rm cool}^{\rm obs} \geq$1\,s. 
We can express $\gamma$ as a function of the synchrotron frequency,  
$\nu_{\rm s} = 3.6 \times 10^{6}\, B^\prime\,\gamma^{2}\,\Gamma / (1+z)$\,Hz,  
obtaining
\begin{equation}
B^\prime \leq \Biggl[\frac{6\,\pi\,m_{\rm e}\,c\, 
1.9 \times 10^3\,(1+z)^{1/2}}{\sigma_{\rm T}\,\Gamma^{1/2}\,\nu_{\rm s}^{1/2}}\Biggr]^{2/3}~.
\end{equation}
%

%
Considering the typical value of $E_{\rm break}\sim 100$\,keV found in our analysis, we obtain
\begin{equation}
B^\prime \lesssim 13  \,\,\,\,  \Gamma_{2}^{-1/3} \nu_{\rm s,100 \,\rm keV}^{-1/3} \,\,\, \rm Gauss~.
\end{equation}
Such a small value of $B^\prime$ is at odds with the expectations for a dissipation region located at $R\sim10^{13}$--$10^{14}$ cm. Considering a value of $B^{\prime}\sim 10^{15}$ Gauss close to the central powerhouse, conservation of the Poynting flux ($P_{\rm B}\propto R^{2}\Gamma^{2}{B^{\prime}}^{2}$=const) implies $B^{\prime}\sim 10^{5}-10^{6}$ Gauss in the emitting region, not compatible with long cooling timescales   \citep{Ghisellini2000}.

\section{Conclusions}\label{sec:concl}
GRB~160625B is one of the brightest GRBs ever detected by {\it Fermi}-GBM during its nine years of activity.
Its light curve is composed of three distinct emission episode: a precursor, a main event, and a long-lasting, late time, soft emission (see Fig.~\ref{fig:lc}).

We performed time-integrated and time-resolved spectroscopy of the main event, testing different fitting models (see \S~\ref{sec:preliminary}). 
In particular, we introduce a new fitting function, called 2SBPL (Eq.~\ref{eq:2sbpl}), consisting of three smoothly connected power laws. 
Standard models with at most two power laws (e.g. Band and SBPL) fail to give a reasonable fit, both to the time-integrated and time-resolved spectra.
Examples of the fit with a SBPL are shown in Fig.~\ref{fig:avgsp} (top panel) and Fig.~\ref{fig:sp_main} (top panel).

Fitting a 2SBPL model to the data, we obtain well-constrained spectral parameters and  significantly improving fits ($F$-test$>3\sigma$) both for the time-integrated spectrum and for 19 out of the 21 time-resolved spectra. 
The additional PL segment (compared to the Band and SBPL functions) describes the low-energy, hardest  part of the spectrum, connected to the usual peaked function by a break that is quite sharp.

The break energy is around 100\,keV, with little evolution in time.
Moreover, the indices of the power laws below and above the low-energy break are $\langle\alpha_1\rangle= -0.63$ ($\sigma=0.08$) and $\langle\alpha_2\rangle= -1.48$ ($\sigma=0.09$). 
These values are remarkably consistent with those predicted for synchrotron emission from a population of non-thermal electrons.
However, the small ratio between the peak and break energy implies that the electron population does not cool completely, and therefore presents a low-energy cut-off.

In fact, when electrons injected at high energies cool, they produce a power-law distribution $N(\gamma)\propto \gamma^{-2}$ and a corresponding synchrotron spectrum of photon index $\alpha_2= -1.5$.
However,  if the cooling occurring in a dynamical time is incomplete, $N(\gamma)$ will have a low-energy cut-off
at some energy $\gamma_{\rm cool}$, corresponding to a frequency $\nu_{\rm cool}$.
Below $\nu_{\rm cool}$ the synchrotron spectral slope will have a photon index $-2/3$.
We therefore identify $\nu_{\rm cool}$ with the found $E_{\rm break}$.

A 2SBPL, however, is not the only possible model for the observed spectrum.
In fact, the spectral hardening below $E_{\rm break}$ could be produced by adding a BB component to a typical single break spectrum (e.g. SBPL or Band), as can be understood from Fig.~\ref{fig:sketch}. 
On the other hand, this extra BB component must be fine tuned in order to mimic the incomplete cooling case, and this fine tuning must be present in each of the time-resolved spectra we analysed.
Moreover, the detailed analysis of the spectrum at the peak of the light curve, where the 2SBPL model is a preferable fit, gave us arguments in support of the 2SBPL model.
A comparison between the fit probability of the two models in all the time-resolved spectra is shown in Fig.~\ref{fig:prob}: the 2SBPL probability is always higher than or equal to the SBPL+BB probability.

    
Our results suggest that the observed GRB prompt spectrum is due to synchrotron emission. 
If \eb\ corresponds to the $\nu_{\rm cool}$ of the electron population, the implied magnetic field is too small with respect to the typically expected value in the emission region, as discussed in \S~\ref{sec:disc}. These results  suggest that further investigation and a revision of the  standard prompt emission model seem necessary.

\begin{acknowledgements}
We thank the referee for the useful comments. M.\,E.\,R. is thankful to the Observatory of Brera for the kind hospitality. 
L.N. acknowledges funding from the European Union's Horizon 2020 Research and Innovation programme under the Marie Sk\l odowska-Curie grant agreement n.\,664931.
We are very grateful to O. Salafia for generous technical support during this work. 
This research has made use of data obtained through the High Energy Astrophysics Science Archive Research Center Online Service, provided by the NASA/Goddard Space Flight Center, and specifically this work made use of public {\it Fermi}--GBM data. 
\end{acknowledgements}

\bibliographystyle{aa} 
\bibliography{references} 

\begin{thebibliography}{38}
\expandafter\ifx\csname natexlab\endcsname\relax\def\natexlab#1{#1}\fi

\bibitem[{{Alexander} {et~al.}(2017){Alexander}, {Laskar}, {Berger},
  {Guidorzi}, {Dichiara}, {Fong}, {Gomboc}, {Kobayashi}, {Kopac}, {Mundell},
  {Tanvir}, \& {Williams}}]{Alexander2017}
{Alexander}, K.~D., {Laskar}, T., {Berger}, E., {et~al.} 2017, \apj, 848, 69

\bibitem[{{Band} {et~al.}(1993){Band}, {Matteson}, {Ford}, {Schaefer},
  {Palmer}, {Teegarden}, {Cline}, {Briggs}, {Paciesas}, {Pendleton}, {Fishman},
  {Kouveliotou}, {Meegan}, {Wilson}, \& {Lestrade}}]{Band1993}
{Band}, D., {Matteson}, J., {Ford}, L., {et~al.} 1993, \apj, 413, 281

\bibitem[{{Beniamini} \& {Piran}(2013)}]{Beniamini2013}
{Beniamini}, P. \& {Piran}, T. 2013, \apj, 769, 69

\bibitem[{{Cabrera} {et~al.}(2007){Cabrera}, {Firmani}, {Avila-Reese},
  {Ghirlanda}, {Ghisellini}, \& {Nava}}]{Cabrera2007}
{Cabrera}, J.~I., {Firmani}, C., {Avila-Reese}, V., {et~al.} 2007, \mnras, 382,
  342

\bibitem[{{Calderone} {et~al.}(2015){Calderone}, {Ghirlanda}, {Ghisellini},
  {Bernardini}, {Campana}, {Covino}, {D'Avanzo}, {Melandri}, {Salvaterra},
  {Sbarufatti}, \& {Tagliaferri}}]{Calderone2015}
{Calderone}, G., {Ghirlanda}, G., {Ghisellini}, G., {et~al.} 2015, \mnras, 448,
  403

\bibitem[{{Daigne} {et~al.}(2011){Daigne}, {Bo{\v s}njak}, \&
  {Dubus}}]{Daigne11}
{Daigne}, F., {Bo{\v s}njak}, {\v Z}., \& {Dubus}, G. 2011, \aap, 526, A110

\bibitem[{{Derishev} {et~al.}(2001){Derishev}, {Kocharovsky}, \&
  {Kocharovsky}}]{Derishev01}
{Derishev}, E.~V., {Kocharovsky}, V.~V., \& {Kocharovsky}, V.~V. 2001, \aap,
  372, 1071

\bibitem[{{Frontera} {et~al.}(2000){Frontera}, {Amati}, {Costa}, {Muller},
  {Pian}, {Piro}, {Soffitta}, {Tavani}, {Castro-Tirado}, {Dal Fiume}, {Feroci},
  {Heise}, {Masetti}, {Nicastro}, {Orlandini}, {Palazzi}, \&
  {Sari}}]{Frontera2000}
{Frontera}, F., {Amati}, L., {Costa}, E., {et~al.} 2000, \apjs, 127, 59

\bibitem[{{Ghirlanda} {et~al.}(2002){Ghirlanda}, {Celotti}, \&
  {Ghisellini}}]{Ghirlanda2002}
{Ghirlanda}, G., {Celotti}, A., \& {Ghisellini}, G. 2002, \aap, 393, 409

\bibitem[{{Ghisellini} {et~al.}(2000){Ghisellini}, {Celotti}, \&
  {Lazzati}}]{Ghisellini2000}
{Ghisellini}, G., {Celotti}, A., \& {Lazzati}, D. 2000, \mnras, 313, L1

\bibitem[{{Goldstein} {et~al.}(2012){Goldstein}, {Burgess}, {Preece}, {Briggs},
  {Guiriec}, {van der Horst}, {Connaughton}, {Wilson-Hodge}, {Paciesas},
  {Meegan}, {von Kienlin}, {Bhat}, {Bissaldi}, {Chaplin}, {Diehl}, {Fishman},
  {Fitzpatrick}, {Foley}, {Gibby}, {Giles}, {Greiner}, {Gruber}, {Kippen},
  {Kouveliotou}, {McBreen}, {McGlynn}, {Rau}, \& {Tierney}}]{Goldstein12}
{Goldstein}, A., {Burgess}, J.~M., {Preece}, R.~D., {et~al.} 2012, \apjs, 199,
  19

\bibitem[{{Gruber} {et~al.}(2014){Gruber}, {Goldstein}, {Weller von Ahlefeld},
  {Narayana Bhat}, {Bissaldi}, {Briggs}, {Byrne}, {Cleveland}, {Connaughton},
  {Diehl}, {Fishman}, {Fitzpatrick}, {Foley}, {Gibby}, {Giles}, {Greiner},
  {Guiriec}, {van der Horst}, {von Kienlin}, {Kouveliotou}, {Layden}, {Lin},
  {Meegan}, {McGlynn}, {Paciesas}, {Pelassa}, {Preece}, {Rau}, {Wilson-Hodge},
  {Xiong}, {Younes}, \& {Yu}}]{Gruber2014}
{Gruber}, D., {Goldstein}, A., {Weller von Ahlefeld}, V., {et~al.} 2014, \apjs,
  211, 12

\bibitem[{{Kaneko} {et~al.}(2006){Kaneko}, {Preece}, {Briggs}, {Paciesas},
  {Meegan}, \& {Band}}]{Kaneko2006}
{Kaneko}, Y., {Preece}, R.~D., {Briggs}, M.~S., {et~al.} 2006, \apjs, 166, 298

\bibitem[{{Katz}(1994)}]{Katz94}
{Katz}, J.~I. 1994, \apjl, 432, L107

\bibitem[{{Lien} {et~al.}(2016){Lien}, {Sakamoto}, {Barthelmy}, {Baumgartner},
  {Cannizzo}, {Chen}, {Collins}, {Cummings}, {Gehrels}, {Krimm}, {Markwardt},
  {Palmer}, {Stamatikos}, {Troja}, \& {Ukwatta}}]{Lien2016}
{Lien}, A., {Sakamoto}, T., {Barthelmy}, S.~D., {et~al.} 2016, \apj, 829, 7

\bibitem[{{Lithwick} \& {Sari}(2001)}]{lithwick01}
{Lithwick}, Y. \& {Sari}, R. 2001, \apj, 555, 540

\bibitem[{{Lloyd} \& {Petrosian}(2000)}]{Lloyd2000}
{Lloyd}, N.~M. \& {Petrosian}, V. 2000, \apj, 543, 722

\bibitem[{{L{\"u}} {et~al.}(2017){L{\"u}}, {L{\"u}}, {Zhong}, {Huang}, {Zhang},
  {Xie}, {Lu}, \& {Liang}}]{Lu2017}
{L{\"u}}, H.-J., {L{\"u}}, J., {Zhong}, S.-Q., {et~al.} 2017, ArXiv e-prints
  [\eprint[arXiv]{1702.01382}]

\bibitem[{{Medvedev}(2000)}]{Medvedev2000}
{Medvedev}, M.~V. 2000, \apj, 540, 704

\bibitem[{{Nakar} {et~al.}(2009){Nakar}, {Ando}, \& {Sari}}]{Nakar09}
{Nakar}, E., {Ando}, S., \& {Sari}, R. 2009, \apj, 703, 675

\bibitem[{{Nava} {et~al.}(2011){Nava}, {Ghirlanda}, {Ghisellini}, \&
  {Celotti}}]{Nava2011}
{Nava}, L., {Ghirlanda}, G., {Ghisellini}, G., \& {Celotti}, A. 2011, \mnras,
  415, 3153

\bibitem[{{Oganesyan} {et~al.}(2017{\natexlab{a}}){Oganesyan}, {Nava},
  {Ghirlanda}, \& {Celotti}}]{gor2017a}
{Oganesyan}, G., {Nava}, L., {Ghirlanda}, G., \& {Celotti}, A.
  2017{\natexlab{a}}, \apj, 846, 137

\bibitem[{{Oganesyan} {et~al.}(2017{\natexlab{b}}){Oganesyan}, {Nava},
  {Ghirlanda}, \& {Celotti}}]{gor2017b}
{Oganesyan}, G., {Nava}, L., {Ghirlanda}, G., \& {Celotti}, A.
  2017{\natexlab{b}}, ArXiv e-prints [\eprint[arXiv]{1710.09383}]

\bibitem[{{Pe'er} \& {Zhang}(2006)}]{Peer2006}
{Pe'er}, A. \& {Zhang}, B. 2006, \apj, 653, 454

\bibitem[{{Preece} {et~al.}(1998){Preece}, {Briggs}, {Mallozzi}, {Pendleton},
  {Paciesas}, \& {Band}}]{Preece1998}
{Preece}, R.~D., {Briggs}, M.~S., {Mallozzi}, R.~S., {et~al.} 1998, \apjl, 506,
  L23

\bibitem[{{Protassov} {et~al.}(2002){Protassov}, {van Dyk}, {Connors},
  {Kashyap}, \& {Siemiginowska}}]{Protassov2002}
{Protassov}, R., {van Dyk}, D.~A., {Connors}, A., {Kashyap}, V.~L., \&
  {Siemiginowska}, A. 2002, \apj, 571, 545

\bibitem[{{Rees} \& {Meszaros}(1994)}]{Rees94}
{Rees}, M.~J. \& {Meszaros}, P. 1994, \apjl, 430, L93

\bibitem[{{Sakamoto} {et~al.}(2011){Sakamoto}, {Barthelmy}, {Baumgartner},
  {Cummings}, {Fenimore}, {Gehrels}, {Krimm}, {Markwardt}, {Palmer}, {Parsons},
  {Sato}, {Stamatikos}, {Tueller}, {Ukwatta}, \& {Zhang}}]{Sakamoto2011}
{Sakamoto}, T., {Barthelmy}, S.~D., {Baumgartner}, W.~H., {et~al.} 2011, \apjs,
  195, 2

\bibitem[{{Sari} {et~al.}(1996){Sari}, {Narayan}, \& {Piran}}]{Sari96}
{Sari}, R., {Narayan}, R., \& {Piran}, T. 1996, \apj, 473, 204

\bibitem[{{Sari} {et~al.}(1998){Sari}, {Piran}, \& {Narayan}}]{Sari98}
{Sari}, R., {Piran}, T., \& {Narayan}, R. 1998, \apjl, 497, L17

\bibitem[{{Tavani}(1996)}]{Tavani96}
{Tavani}, M. 1996, \apj, 466, 768

\bibitem[{Troja {et~al.}(2017)Troja, Lipunov, Mundell, Butler, Watson,
  Kobayashi, Cenko, Marshall, Ricci, Fruchter, Wieringa, Gorbovskoy, Kornilov,
  Kutyrev, Lee, Toy, Tyurina, Budnev, Buckley, Gonz{\'a}lez, Gress, Horesh,
  Panasyuk, Prochaska, Ramirez-Ruiz, Lopez, Richer, Rom{\'a}n-Z{\'u}{\~n}iga,
  Serra-Ricart, Yurkov, \& Gehrels}]{Troja2017}
Troja, E., Lipunov, V.~M., Mundell, C.~G., {et~al.} 2017, Nature, 547, 425,
  letter

\bibitem[{{Uhm} \& {Zhang}(2014)}]{Uhm2014}
{Uhm}, Z.~L. \& {Zhang}, B. 2014, Nature Physics, 10, 351

\bibitem[{{Vianello} {et~al.}(2017){Vianello}, {Gill}, {Granot}, {Omodei},
  {Cohen-Tanugi}, \& {Longo}}]{vianello17}
{Vianello}, G., {Gill}, R., {Granot}, J., {et~al.} 2017, ArXiv e-prints
  [\eprint[arXiv]{1706.01481}]

\bibitem[{{Wang} {et~al.}(2017){Wang}, {Wang}, {Zhang}, {Liang}, {Jin}, {He},
  {Liao}, {Fan}, \& {Wei}}]{Wang2017}
{Wang}, Y.-Z., {Wang}, H., {Zhang}, S., {et~al.} 2017, \apj, 836, 81

\bibitem[{{Xu} {et~al.}(2016){Xu}, {Malesani}, {Fynbo}, {Tanvir}, {Levan}, \&
  {Perley}}]{xu16}
{Xu}, D., {Malesani}, D., {Fynbo}, J.~P.~U., {et~al.} 2016, GRB Coordinates
  Network, Circular Service, No.~19600, \#1 (2016), 9600

\bibitem[{{Zhang} {et~al.}(2016){Zhang}, {Zhang}, {Castro-Tirado}, {Dai},
  {Tam}, {Wang}, {Hu}, {Karpov}, {Pozanenko}, {Zhang}, {Mazaeva}, {Minaev},
  {Volnova}, {Oates}, {Gao}, {Wu}, {Shao}, {Tang}, {Beskin}, {Biryukov},
  {Bondar}, {Ivanov}, {Katkova}, {Orekhova}, {Perkov}, {Sasyuk}, {Mankiewicz},
  {{\.Z}arnecki}, {Cwiek}, {Opiela}, {Zadro{\.z}ny}, {Aptekar}, {Frederiks},
  {Svinkin}, {Kusakin}, {Inasaridze}, {Burhonov}, {Rumyantsev}, {Klunko},
  {Moskvitin}, {Fatkhullin}, {Sokolov}, {Valeev}, {Jeong}, {Park},
  {Caballero-Garc{\'{\i}}a}, {Cunniffe}, {Tello}, {Ferrero}, {Pandey},
  {Jel{\'{\i}}nek}, {S{\'a}nchez-Ram{\'{\i}}rez}, \&
  {Castell{\'o}n}}]{Zhang2016}
{Zhang}, B.-B., {Zhang}, B., {Castro-Tirado}, A.~J., {et~al.} 2016, ArXiv
  e-prints [\eprint[arXiv]{1612.03089}]

\bibitem[{{Zhao} {et~al.}(2014){Zhao}, {Li}, {Liu}, {Zhang}, {Bai}, \&
  {M{\'e}sz{\'a}ros}}]{Zhao2014}
{Zhao}, X., {Li}, Z., {Liu}, X., {et~al.} 2014, \apj, 780, 12

\end{thebibliography}

\newpage
\begin{appendix}

\section{Precursor and last emission episode}\label{app:prec_and_last}

GRB~160625B is characterized by three well separated emission episodes, as shown in the light curve of Fig.~\ref{fig:lc}.
The first (precursor) and last episode are visible only in the NaI detectors. 
While in the main text we presented the results of the analysis of the main event, here we report on the analysis of the precursor (that triggered the GBM) and of the last emission episode. 
Since the emission is much fainter than the main event, we perform only time-integrated analysis for both the precursor and the last event.

\begin{figure}[hb]
\centering
\includegraphics[scale=0.14]{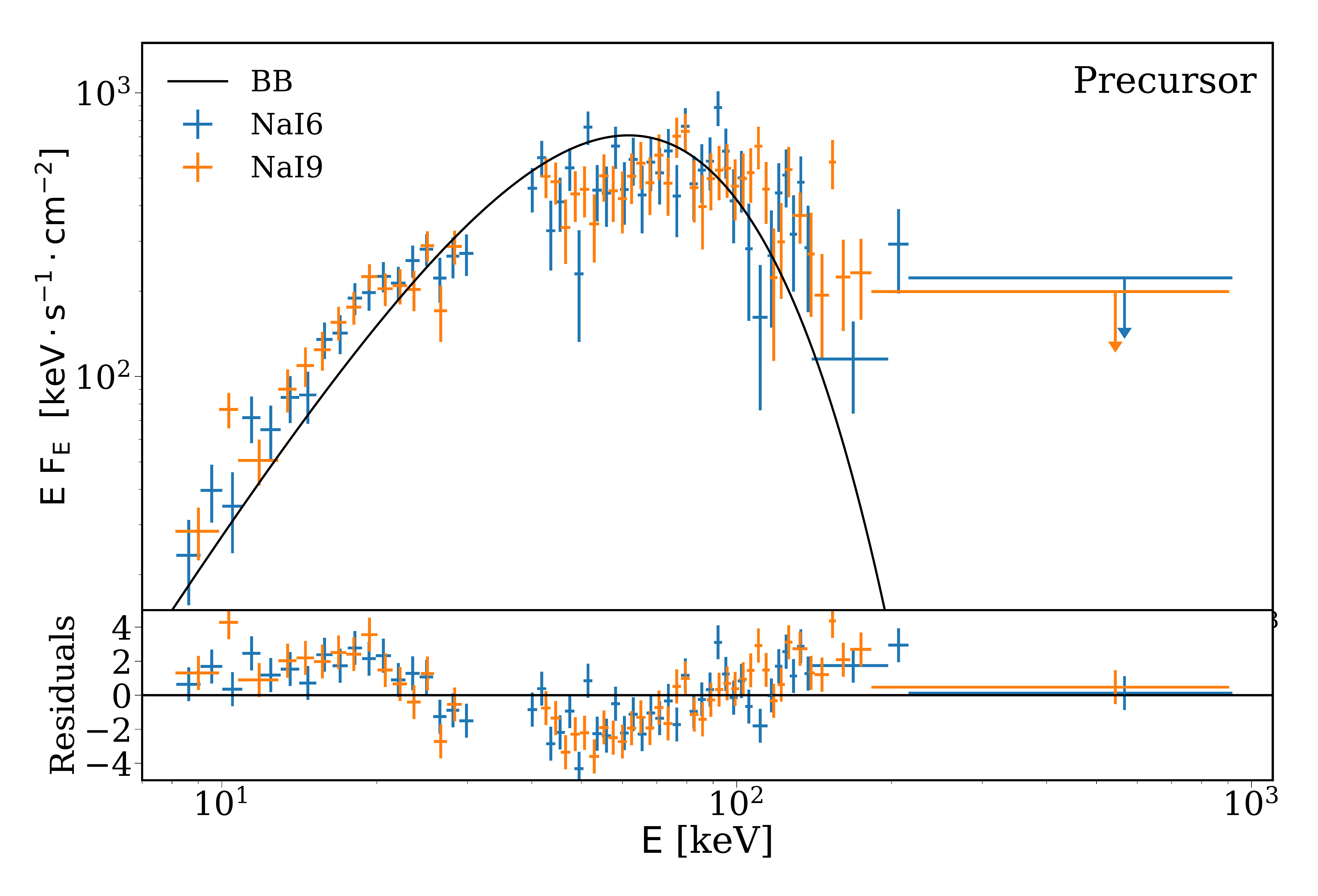}
\includegraphics[scale=0.14]{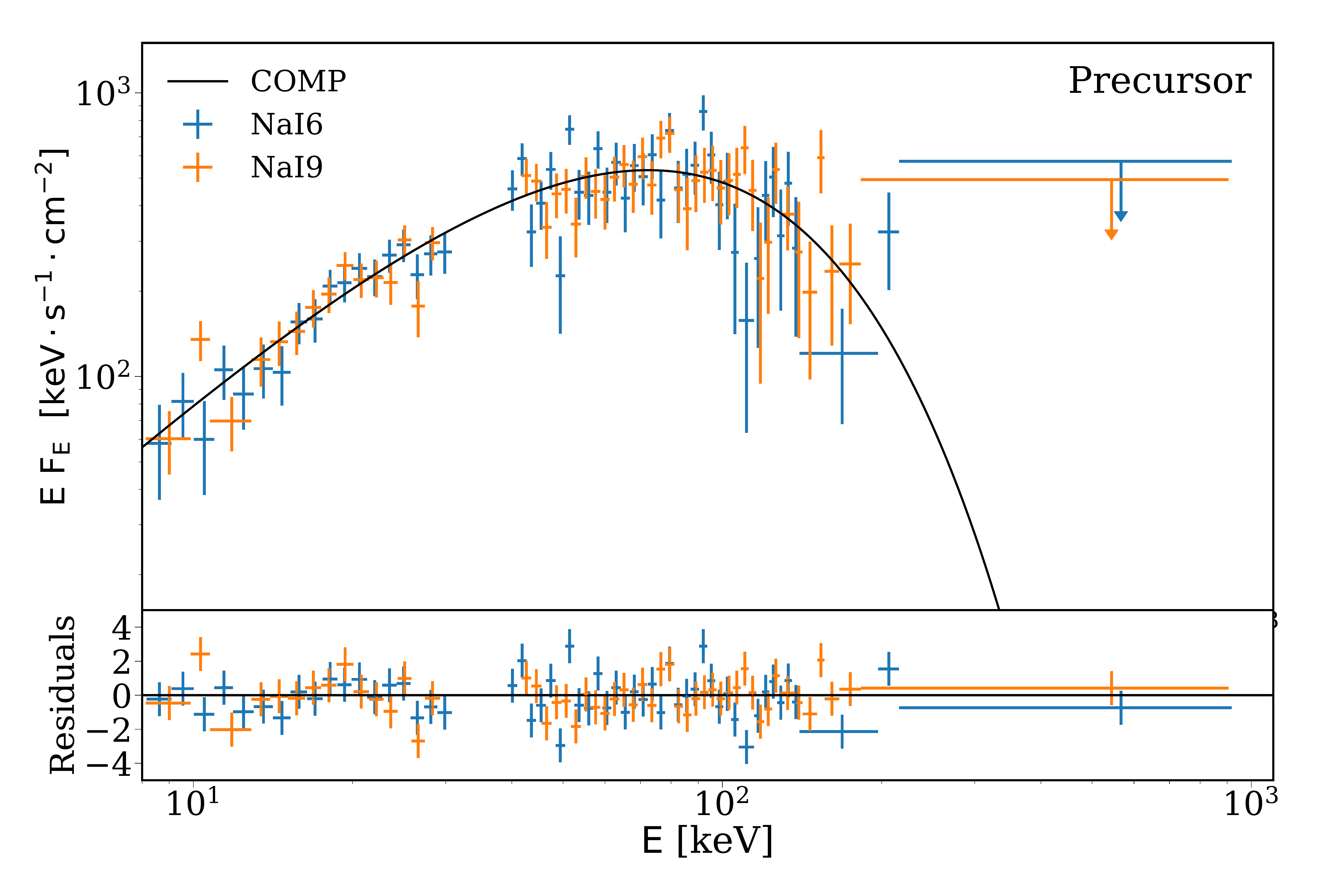}
\caption{Top panel: $EF_{\rm E}$ spectrum of the precursor (0.002--1.056\,s) fitted with a single BB (dotted line).  Error bars on the data points denote the 1$\sigma$ confidence level. The 3$\sigma$ upper limits are shown with arrows. Residuals in units of $\sigma$ are shown in the bottom stripe. Bottom panel: Same spectrum fitted with the CPL model.}  \label{fig:prec}
\end{figure}

We fit the spectrum of the precursor from 0.002\,s to 1.056\,s with two models: a BB (top panel in Fig.~\ref{fig:prec}) and a CPL (bottom panel). 
The BGO data are not included in the analysis.
The fit with a CPL model improves the fit over the single BB (at odds with respect to what has been found by \cite{Zhang2016}), with a reduction of the Cash statistics\footnote{Due to the reduced photon flux, these spectra are fitted with the Cash statistics.} of 214. For the BB model, we obtain kT = $15.47_{-0.24}^{+0.25}$ keV and an amplitude parameter A = $2.46_{-0.15}^{+0.16} \cdot 10^{-3} \,\, \rm ph \, s^{-1} \, cm^{-2} \, keV^{-1}$, with $\rm Cstat_{BB} = 495.44$.
For the CPL model, we obtain the slope of the power law $\alpha = -0.21\pm-0.09$, a peak energy $E_{\rm peak}=70.05_{-1.85}^{+1.98}$ keV, and an amplitude A = $0.60_{-0.10}^{0.12} \,\, \rm ph \, s^{-1} \, cm^{-2} \, keV^{-1}$, with a $\rm Cstat_{COMP} = 281.41$.
As can be seen from  the residuals of the BB model, this function is too narrow to account for the observed spectrum.
To test whether this can be caused by the temporal evolution of a BB spectrum, we split the time interval of the precursor (1.054 seconds) in three sub-intervals and perform the analysis in each of them.
Again, we find that the fit with a CPL model is better than the BB. 

\begin{figure}
\centering
\includegraphics[scale=0.14]{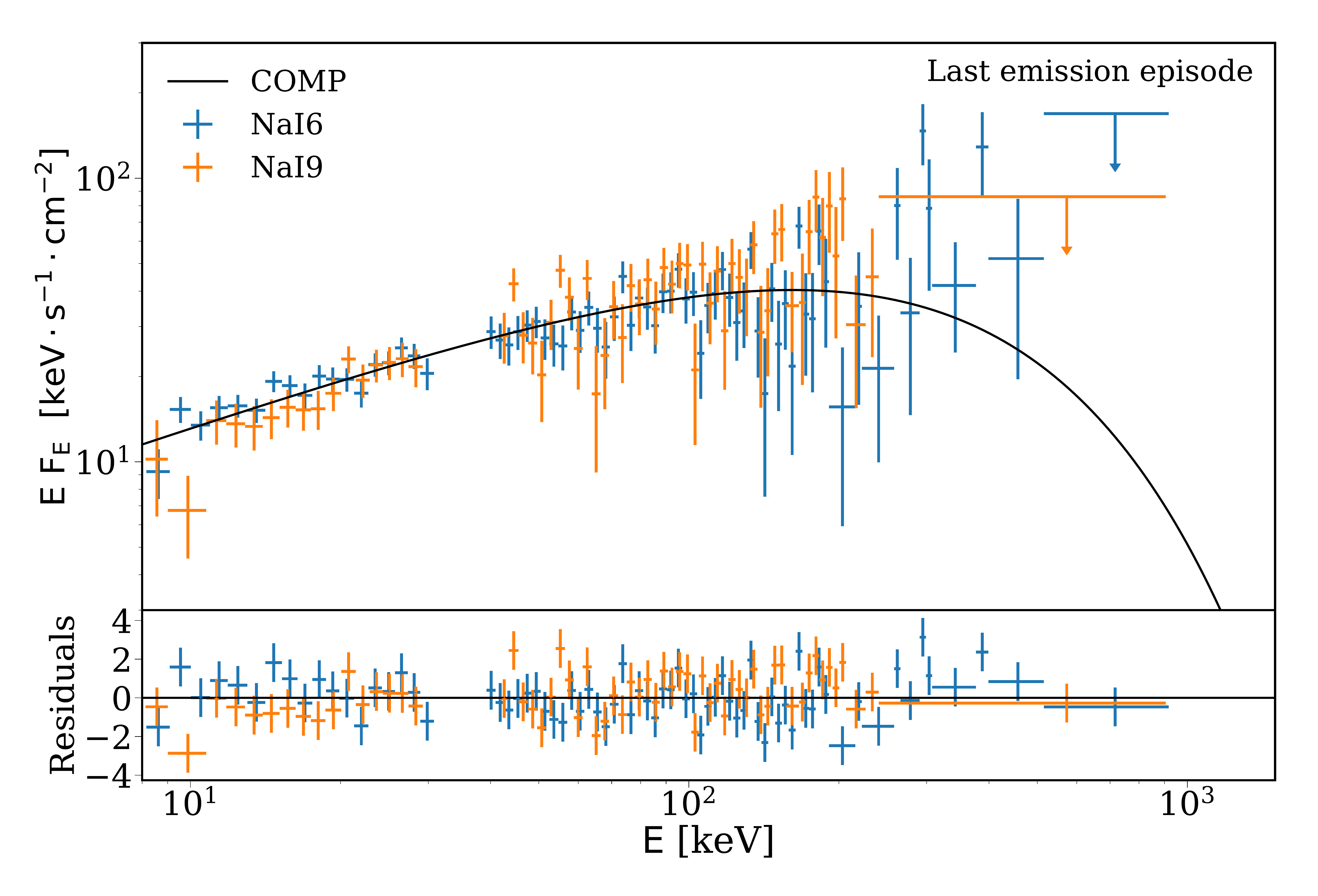}
\caption{Spectrum of the last emission episode (281 seconds in width) of the GRB~160625B fitted with the CPL model.}  \label{fig:lastepisode}
\end{figure}

The last emission episode consists of a smooth and dim pulse of long duration.
In the time interval 519.21-- 800.05 \,s, the GBM spectrum is best fitted by a CPL model, as shown in Fig.~\ref{fig:lastepisode}. We obtain the slope of the power law $\alpha = -1.50\pm-0.03$ , a peak energy $E_{\rm peak}=135.4_{-9.4}^{+11.2}$ keV, and an amplitude A = $4.5\pm0.2 \cdot 10^{-3} \,\, \rm ph \, s^{-1} \, cm^{-2} \, keV^{-1}$, with a $\rm Cstat_{COMP} = 518.03$.    

In this time interval the LAT emission becomes relevant \citep[e.g.]{Troja2017}.
We fit the LAT data alone and the spectrum is best fitted by a power law with spectral index $\gamma = -1.98 \pm 0.22$ and normalization $N = (1.61 \pm 0.43)\times 10^{-5}$ph\,cm$^{-2}$\,s$^{-1}$.
The LAT spectrum is then inconsistent with the extrapolation of the GBM spectrum to the highest energies.

\section{Tables}\label{app:tables}
\begin{table*}
\centering 
\caption{Best fit parameters for the SBPL model for the main emission episode of GRB~160625B.}
\label{tab:sbpl}
\scriptsize
\begin{tabular}{ccccccccc}
\hline\hline
 \multicolumn{1}{c}{Time Bin} &
 \multicolumn{1}{c}{Norm.} &
 \multicolumn{1}{c}{$\alpha$} &
 \multicolumn{1}{c}{$E_{peak}$} &
 \multicolumn{1}{c}{$\beta$} &
 \multicolumn{1}{c}{Photon Flux} &
 \multicolumn{1}{c}{Energy Flux $(10^{-5})$} &
 \multicolumn{1}{c}{$\chi^{2}$/DOF} &
 \multicolumn{1}{c}{Prob} \\
 $[s]$ &  &  & $[keV]$ & & $[ph \, s^{-1} \, cm^{-2}]$ & $[erg \, s^{-1} \, cm^{-2}]$ &  &  \\
\hline
186.40 - 187.43 & ${ 2.16 }_{- 0.44 }^{+ 0.4 }$ & ${ -0.84 }_{- 0.06 }^{+ 0.04 }$ & ${ 357.0 }_{- 84.1 }^{+ 81.9 }$ & ${ -1.81 }_{- 0.07 }^{+ 0.03 }$ & $ 20.29 \pm 0.44 $ & $ 2.6 \pm 0.2 $ & 464.3 / 464 & 0.49 \\
187.43 - 188.45 & ${ 4.19 }_{- 0.41 }^{+ 0.39 }$ & ${ -0.67 }_{- 0.03 }^{+ 0.02 }$ & ${ 278.5 }_{- 21.2 }^{+ 20.1 }$ & ${ -1.9 }_{- 0.02 }^{+ 0.02 }$ & $ 79.93 \pm 0.68 $ & $ 8.5 \pm 0.2 $ & 634.3 / 464 & $6.3 \cdot 10^{-8}$ \\
188.45 - 189.47 & ${ 8.93 }_{- 0.54 }^{+ 0.5 }$ & ${ -0.6 }_{- 0.02 }^{+ 0.01 }$ & ${ 230.8 }_{- 8.47 }^{+ 7.62 }$ & ${ -1.98 }_{- 0.01 }^{+ 0.01 }$ & $ 213.0 \pm 1.1 $ & $ 17.6 \pm 0.2 $ & 1053.6 / 464 & $8.13 \cdot 10^{-48}$ \\
189.47 - 190.50 & ${ 9.91 }_{- 0.58 }^{+ 0.6 }$ & ${ -0.65 }_{- 0.02 }^{+ 0.01 }$ & ${ 456.5 }_{- 22.4 }^{+ 29.4 }$ & ${ -2.06 }_{- 0.01 }^{+ 0.01 }$ & $ 172.84 \pm 1.0 $ & $ 10.7 \pm 0.2 $ & 819.3 / 464 & $4.6 \cdot 10^{-22}$ \\
190.50 - 191.52 & ${ 7.98 }_{- 0.65 }^{+ 0.55 }$ & ${ -0.7 }_{- 0.02 }^{+ 0.02 }$ & ${ 282.3 }_{- 9.2 }^{+ 13.7 }$ & ${ -2.17 }_{- 0.03 }^{+ 0.02 }$ & $ 98.33 \pm 0.78 $ & $ 4.2 \pm 0.1 $ & 543.3 / 464 & 0.01 \\
191.52 - 192.55 & ${ 8.63 }_{- 0.8 }^{+ 0.65 }$ & ${ -0.76 }_{- 0.03 }^{+ 0.02 }$ & ${ 229.2 }_{- 7.33 }^{+ 12.7 }$ & ${ -2.22 }_{- 0.05 }^{+ 0.02 }$ & $ 77.82 \pm 0.72 $ & $ 2.7 \pm 0.1 $ & 458.4 / 464 & 0.56 \\
192.55 - 193.57 & ${ 8.33 }_{- 0.64 }^{+ 0.59 }$ & ${ -0.71 }_{- 0.02 }^{+ 0.02 }$ & ${ 235.1 }_{- 6.54 }^{+ 8.24 }$ & ${ -2.29 }_{- 0.04 }^{+ 0.02 }$ & $ 92.95 \pm 0.77 $ & $ 3.2 \pm 0.1 $ & 499.6 / 464 & 0.12 \\
193.57 - 194.59 & ${ 9.29 }_{- 0.57 }^{+ 0.66 }$ & ${ -0.7 }_{- 0.02 }^{+ 0.02 }$ & ${ 300.0 }_{- 7.44 }^{+ 11.2 }$ & ${ -2.25 }_{- 0.02 }^{+ 0.02 }$ & $ 122.75 \pm 0.86 $ & $ 5.4 \pm 0.1 $ & 644.3 / 464 & $5.5 \cdot 10^{-8}$ \\
194.59 - 195.62 & ${ 10.56 }_{- 0.67 }^{+ 0.64 }$ & ${ -0.7 }_{- 0.02 }^{+ 0.02 }$ & ${ 293.6 }_{- 7.33 }^{+ 9.6 }$ & ${ -2.26 }_{- 0.03 }^{+ 0.01 }$ & $ 135.13 \pm 0.9 $ & $ 5.7 \pm 0.1 $ & 694.9 / 464 & $1.8 \cdot 10^{-11}$ \\
195.62 - 196.64 & ${ 9.46 }_{- 0.66 }^{+ 0.58 }$ & ${ -0.71 }_{- 0.02 }^{+ 0.02 }$ & ${ 279.6 }_{- 7.63 }^{+ 8.78 }$ & ${ -2.29 }_{- 0.03 }^{+ 0.02 }$ & $ 117.21 \pm 0.84 $ & $ 4.7 \pm 0.1 $ & 644.0 / 464 & $5.7 \cdot 10^{-8}$ \\
196.64 - 197.67 & ${ 10.67 }_{- 0.85 }^{+ 0.69 }$ & ${ -0.79 }_{- 0.02 }^{+ 0.02 }$ & ${ 287.9 }_{- 9.77 }^{+ 11.1 }$ & ${ -2.32 }_{- 0.05 }^{+ 0.02 }$ & $ 94.06 \pm 0.79 $ & $ 3.5 \pm 0.1 $ & 570.3 / 464 & $5 \cdot 10^{-4}$ \\
197.67 - 198.69 & ${ 10.21 }_{- 0.7 }^{+ 0.75 }$ & ${ -0.77 }_{- 0.02 }^{+ 0.02 }$ & ${ 316.8 }_{- 9.64 }^{+ 14.4 }$ & ${ -2.25 }_{- 0.03 }^{+ 0.02 }$ & $ 99.4 \pm 0.8 $ & $ 4.2 \pm 0.1 $ & 565.2 / 464 & $9 \cdot 10^{-4}$ \\
198.69 - 199.71 & ${ 9.61 }_{- 0.63 }^{+ 0.69 }$ & ${ -0.76 }_{- 0.02 }^{+ 0.02 }$ & ${ 323.0 }_{- 9.26 }^{+ 13.3 }$ & ${ -2.34 }_{- 0.04 }^{+ 0.02 }$ & $ 103.34 \pm 0.8 $ & $ 4.3 \pm 0.1 $ & 654.9 / 464 & $1.1 \cdot 10^{-8}$ \\
199.71 - 200.74 & ${ 10.44 }_{- 0.64 }^{+ 0.63 }$ & ${ -0.72 }_{- 0.02 }^{+ 0.01 }$ & ${ 472.3 }_{- 16.7 }^{+ 18.5 }$ & ${ -2.13 }_{- 0.02 }^{+ 0.01 }$ & $ 145.68 \pm 0.92 $ & $ 9.1 \pm 0.2 $ & 688.0 / 464 & $5.9 \cdot 10^{-11}$ \\
200.74 - 201.76 & ${ 9.74 }_{- 0.63 }^{+ 0.6 }$ & ${ -0.7 }_{- 0.02 }^{+ 0.02 }$ & ${ 368.6 }_{- 11.8 }^{+ 14.1 }$ & ${ -2.16 }_{- 0.02 }^{+ 0.01 }$ & $ 133.83 \pm 0.9 $ & $ 7.0 \pm 0.1 $ & 606.6 / 464 & $8.8 \cdot 10^{-6}$ \\
201.76 - 202.79 & ${ 12.89 }_{- 0.73 }^{+ 0.94 }$ & ${ -0.79 }_{- 0.01 }^{+ 0.02 }$ & ${ 324.9 }_{- 8.65 }^{+ 14.2 }$ & ${ -2.35 }_{- 0.03 }^{+ 0.02 }$ & $ 118.86 \pm 0.86 $ & $ 4.8 \pm 0.1 $ & 653.5 / 464 & $1.4 \cdot 10^{-8}$ \\
202.79 - 203.81 & ${ 12.37 }_{- 1.01 }^{+ 0.76 }$ & ${ -0.86 }_{- 0.02 }^{+ 0.02 }$ & ${ 297.4 }_{- 11.0 }^{+ 13.0 }$ & ${ -2.35 }_{- 0.06 }^{+ 0.02 }$ & $ 83.57 \pm 0.74 $ & $ 3 \pm 0.1 $ & 530.4 / 464 & 0.02 \\
203.81 - 204.83 & ${ 9.67 }_{- 0.79 }^{+ 0.8 }$ & ${ -0.85 }_{- 0.02 }^{+ 0.02 }$ & ${ 323.1 }_{- 12.3 }^{+ 21.3 }$ & ${ -2.24 }_{- 0.05 }^{+ 0.03 }$ & $ 67.53 \pm 0.68 $ & $ 2.7 \pm 0.1 $ & 504.3 / 464 & 0.1 \\
204.83 - 205.86 & ${ 8.33 }_{- 0.59 }^{+ 0.68 }$ & ${ -0.81 }_{- 0.02 }^{+ 0.02 }$ & ${ 261.8 }_{- 8.74 }^{+ 11.4 }$ & ${ -2.32 }_{- 0.04 }^{+ 0.03 }$ & $ 64.98 \pm 0.67 $ & $ 2.2 \pm 0.1 $ & 511.7 / 464 & 0.06 \\
205.86 - 206.88 & ${ 6.63 }_{- 0.77 }^{+ 0.6 }$ & ${ -0.78 }_{- 0.03 }^{+ 0.02 }$ & ${ 203.4 }_{- 7.01 }^{+ 11.6 }$ & ${ -2.31 }_{- 0.07 }^{+ 0.03 }$ & $ 54.05 \pm 0.64 $ & $ 1.6 \pm 0.1 $ & 480.4 / 464 & 0.29 \\
206.88 - 207.91 & ${ 6.74 }_{- 0.92 }^{+ 0.71 }$ & ${ -0.84 }_{- 0.04 }^{+ 0.03 }$ & ${ 253.0 }_{- 14.6 }^{+ 81.6 }$ & ${ -2.13 }_{- 0.09 }^{+ 0.03 }$ & $ 42.8 \pm 0.58 $ & $ 1.5 \pm 0.1 $ & 508.3 / 464 & 0.08 \\
\hline
\end{tabular}
\end{table*}

\begin{table*}
\centering
\caption{Best fit parameters for the 2SBPL model for the main emission episode of GRB160625B.}
\label{tab:2sbpl}
\scriptsize
\begin{tabular}{cccccccccccc}
\hline\hline
  \multicolumn{1}{c}{Time Bin} &
  \multicolumn{1}{c}{Norm.} &
  \multicolumn{1}{c}{$\alpha_{1}$} &
  \multicolumn{1}{c}{$E_{break}$} &
  \multicolumn{1}{c}{$\alpha_{2}$} &
  \multicolumn{1}{c}{$E_{peak}$} &
  \multicolumn{1}{c}{$\beta$} &
  \multicolumn{1}{c}{Photon Flux} &
  \multicolumn{1}{c}{Energy Flux $(10^{-5})$} &
  \multicolumn{1}{c}{$\chi^{2}$/DOF} &
  \multicolumn{1}{c}{Prob.} &
  \multicolumn{1}{c}{$\sigma (F_{test})$} \\
   $[s]$ &  &  & $[keV]$ & & $[keV]$ &  & $[ph \, s^{-1} \, cm^{-2}]$ & $[erg \, s^{-1} \, cm^{-2}]$ &  &   & \\
\hline
186.40 - 187.43 & ${ 1.8 }_{- 0.26 }^{+ 0.32 }$ & ${ -0.8 }_{- 0.04 }^{+ 0.04 }$ & ${ 196.7 }_{- 26.3 }^{+ 31.2 }$ & ${ -1.53 }_{- 0.05 }^{+ 0.04 }$ & ${ 6596.0 }_{- 1400.0 }^{+ 1220.0 }$ & ${ -3.88 }_{- 1.03 }^{+ 1.02 }$ & $ 20.16 \pm 0.44 $ & $ 2.3 \pm 0.1 $ & 438.2 / 462 & 0.78 & 4.8 \\
187.43 - 188.45 & ${ 2.68 }_{- 0.31 }^{+ 0.32 }$ & ${ -0.55 }_{- 0.03 }^{+ 0.03 }$ & ${ 138.1 }_{- 12.1 }^{+ 11.5 }$ & ${ -1.45 }_{- 0.05 }^{+ 0.04 }$ & ${ 2458.0 }_{- 248.0 }^{+ 258.0 }$ & ${ -2.4 }_{- 0.09 }^{+ 0.09 }$ & $ 79.58 \pm 0.69 $ & $ 7.7 \pm 0.2 $ & 487.7 / 462 & 0.2 & $>8.4$ \\
188.45 - 189.47 & ${ 5.9 }_{- 0.37 }^{+ 0.4 }$ & ${ -0.48 }_{- 0.02 }^{+ 0.02 }$ & ${ 127.7 }_{- 4.92 }^{+ 5.25 }$ & ${ -1.49 }_{- 0.03 }^{+ 0.03 }$ & ${ 1547.0 }_{- 73.1 }^{+ 87.2 }$ & ${ -2.54 }_{- 0.05 }^{+ 0.05 }$ & $ 213.67 \pm 1.1 $ & $ 15.8 \pm 0.2 $ & 494.0 / 462 & 0.15 & $>8.4$ \\
189.47 - 190.50 & ${ 6.87 }_{- 0.5 }^{+ 0.49 }$ & ${ -0.54 }_{- 0.02 }^{+ 0.02 }$ & ${ 114.4 }_{- 5.65 }^{+ 5.16 }$ & ${ -1.52 }_{- 0.04 }^{+ 0.03 }$ & ${ 1098.0 }_{- 71.9 }^{+ 68.6 }$ & ${ -2.54 }_{- 0.06 }^{+ 0.04 }$ & $ 173.91 \pm 1.0 $ & $ 10.0 \pm 0.2 $ & 505.7 / 462 & 0.08 & $>8.4$ \\
190.50 - 191.52 & ${ 5.73 }_{- 0.56 }^{+ 0.54 }$ & ${ -0.6 }_{- 0.03 }^{+ 0.02 }$ & ${ 98.17 }_{- 7.22 }^{+ 6.33 }$ & ${ -1.54 }_{- 0.07 }^{+ 0.05 }$ & ${ 621.6 }_{- 54.4 }^{+ 45.8 }$ & ${ -2.54 }_{- 0.09 }^{+ 0.05 }$ & $ 99.09 \pm 0.79 $ & $ 4.0 \pm 0.1 $ & 451.1 / 462 & 0.63 & $>8.4$ \\
191.52 - 192.55 & ${ 7.76 }_{- 0.81 }^{+ 0.72 }$ & ${ -0.73 }_{- 0.03 }^{+ 0.02 }$ & ${ 107.4 }_{- 8.89 }^{+ 6.63 }$ & ${ -1.78 }_{- 0.1 }^{+ 0.05 }$ & ${ 574.6 }_{- 114.0 }^{+ 65.8 }$ & ${ -2.76 }_{- 0.24 }^{+ 0.08 }$ & $ 78.41 \pm 0.73 $ & $ 2.5 \pm 0.1 $ & 422.9 / 462 & 0.9 & 5.8 \\
192.55 - 193.57 & ${ 6.26 }_{- 0.69 }^{+ 0.65 }$ & ${ -0.63 }_{- 0.03 }^{+ 0.03 }$ & ${ 99.2 }_{- 10.2 }^{+ 8.49 }$ & ${ -1.51 }_{- 0.12 }^{+ 0.08 }$ & ${ 387.1 }_{- 39.3 }^{+ 30.8 }$ & ${ -2.53 }_{- 0.09 }^{+ 0.05 }$ & $ 93.29 \pm 0.78 $ & $ 3.1 \pm 0.1 $ & 456.6 / 462 & 0.56 & 6.1 \\
193.57 - 194.59 & ${ 5.97 }_{- 0.55 }^{+ 0.53 }$ & ${ -0.57 }_{- 0.03 }^{+ 0.02 }$ & ${ 102.3 }_{- 7.2 }^{+ 6.25 }$ & ${ -1.48 }_{- 0.05 }^{+ 0.04 }$ & ${ 608.2 }_{- 34.5 }^{+ 26.5 }$ & ${ -2.71 }_{- 0.09 }^{+ 0.04 }$ & $ 123.2 \pm 0.87 $ & $ 5.0 \pm 0.1 $ & 451.8 / 462 & 0.62 & $>8.4$ \\
194.59 - 195.62 & ${ 6.78 }_{- 0.58 }^{+ 0.56 }$ & ${ -0.58 }_{- 0.03 }^{+ 0.02 }$ & ${ 102.0 }_{- 6.67 }^{+ 5.81 }$ & ${ -1.49 }_{- 0.05 }^{+ 0.04 }$ & ${ 590.8 }_{- 33.3 }^{+ 24.1 }$ & ${ -2.72 }_{- 0.08 }^{+ 0.04 }$ & $ 135.62 \pm 0.91 $ & $ 5.3 \pm 0.1 $ & 485.0 / 462 & 0.22 & $>8.4$ \\
195.62 - 196.64 & ${ 6.13 }_{- 0.57 }^{+ 0.54 }$ & ${ -0.58 }_{- 0.03 }^{+ 0.02 }$ & ${ 102.1 }_{- 7.42 }^{+ 6.15 }$ & ${ -1.51 }_{- 0.06 }^{+ 0.04 }$ & ${ 559.0 }_{- 35.4 }^{+ 23.9 }$ & ${ -2.79 }_{- 0.1 }^{+ 0.05 }$ & $ 117.59 \pm 0.85 $ & $ 4.3 \pm 0.1 $ & 474.0 / 462 & 0.34 & $>8.4$ \\
196.64 - 197.67 & ${ 5.76 }_{- 0.9 }^{+ 0.73 }$ & ${ -0.61 }_{- 0.05 }^{+ 0.04 }$ & ${ 83.06 }_{- 11.8 }^{+ 8.38 }$ & ${ -1.42 }_{- 0.08 }^{+ 0.05 }$ & ${ 531.4 }_{- 48.6 }^{+ 28.2 }$ & ${ -2.83 }_{- 0.17 }^{+ 0.08 }$ & $ 94.23 \pm 0.79 $ & $ 3.2 \pm 0.1 $ & 459.4 / 462 & 0.53 & $>8.4$ \\
197.67 - 198.69 & ${ 6.22 }_{- 0.76 }^{+ 0.69 }$ & ${ -0.63 }_{- 0.04 }^{+ 0.03 }$ & ${ 92.24 }_{- 11.0 }^{+ 8.86 }$ & ${ -1.41 }_{- 0.07 }^{+ 0.05 }$ & ${ 570.5 }_{- 37.9 }^{+ 28.9 }$ & ${ -2.62 }_{- 0.1 }^{+ 0.04 }$ & $ 99.66 \pm 0.8 $ & $ 4.0 \pm 0.1 $ & 453.0 / 462 & 0.61 & $>8.4$ \\
198.69 - 199.71 & ${ 5.13 }_{- 0.61 }^{+ 0.55 }$ & ${ -0.58 }_{- 0.04 }^{+ 0.03 }$ & ${ 94.75 }_{- 9.5 }^{+ 7.6 }$ & ${ -1.42 }_{- 0.06 }^{+ 0.04 }$ & ${ 606.4 }_{- 37.7 }^{+ 23.5 }$ & ${ -2.95 }_{- 0.14 }^{+ 0.06 }$ & $ 103.02 \pm 0.81 $ & $ 3.9 \pm 0.1 $ & 468.4 / 462 & 0.41 & $>8.4$ \\
199.71 - 200.74 & ${ 6.19 }_{- 0.55 }^{+ 0.56 }$ & ${ -0.57 }_{- 0.03 }^{+ 0.02 }$ & ${ 108.7 }_{- 8.86 }^{+ 8.69 }$ & ${ -1.34 }_{- 0.05 }^{+ 0.04 }$ & ${ 824.2 }_{- 40.8 }^{+ 41.7 }$ & ${ -2.4 }_{- 0.04 }^{+ 0.03 }$ & $ 145.38 \pm 0.93 $ & $ 8.6 \pm 0.2 $ & 499.0 / 462 & 0.11 & $>8.4$ \\
200.74 - 201.76 & ${ 6.38 }_{- 0.58 }^{+ 0.6 }$ & ${ -0.58 }_{- 0.03 }^{+ 0.02 }$ & ${ 105.4 }_{- 9.08 }^{+ 8.73 }$ & ${ -1.38 }_{- 0.06 }^{+ 0.05 }$ & ${ 660.5 }_{- 37.8 }^{+ 36.5 }$ & ${ -2.39 }_{- 0.04 }^{+ 0.03 }$ & $ 134.32 \pm 0.91 $ & $ 6.8 \pm 0.1 $ & 476.8 / 462 & 0.31 & $>8.4$ \\
201.76 - 202.79 & ${ 7.55 }_{- 0.73 }^{+ 0.68 }$ & ${ -0.64 }_{- 0.03 }^{+ 0.02 }$ & ${ 102.2 }_{- 8.81 }^{+ 7.29 }$ & ${ -1.48 }_{- 0.05 }^{+ 0.04 }$ & ${ 621.9 }_{- 34.4 }^{+ 22.7 }$ & ${ -3.0 }_{- 0.13 }^{+ 0.06 }$ & $ 118.41 \pm 0.86 $ & $ 4.3 \pm 0.1 $ & 423.9 / 462 & 0.9 & $>8.4$ \\
202.79 - 203.81 & ${ 9.27 }_{- 1.07 }^{+ 0.89 }$ & ${ -0.77 }_{- 0.04 }^{+ 0.03 }$ & ${ 113.3 }_{- 17.5 }^{+ 12.4 }$ & ${ -1.51 }_{- 0.12 }^{+ 0.06 }$ & ${ 482.3 }_{- 50.6 }^{+ 27.0 }$ & ${ -2.74 }_{- 0.18 }^{+ 0.05 }$ & $ 83.6 \pm 0.75 $ & $ 2.8 \pm 0.1 $ & 485.3 / 462 & 0.22 & 6.1 \\
203.81 - 204.83 & ${ 6.62 }_{- 1.12 }^{+ 0.87 }$ & ${ -0.74 }_{- 0.06 }^{+ 0.04 }$ & ${ 94.78 }_{- 21.8 }^{+ 14.5 }$ & ${ -1.42 }_{- 0.13 }^{+ 0.07 }$ & ${ 530.4 }_{- 59.2 }^{+ 41.6 }$ & ${ -2.56 }_{- 0.16 }^{+ 0.06 }$ & $ 67.72 \pm 0.69 $ & $ 2.5 \pm 0.1 $ & 468.3 / 462 & 0.41 & 5.5 \\
204.83 - 205.86 & ${ 6.28 }_{- 0.79 }^{+ 0.71 }$ & ${ -0.73 }_{- 0.04 }^{+ 0.03 }$ & ${ 105.1 }_{- 14.0 }^{+ 10.6 }$ & ${ -1.56 }_{- 0.12 }^{+ 0.07 }$ & ${ 463.8 }_{- 55.7 }^{+ 32.5 }$ & ${ -2.83 }_{- 0.23 }^{+ 0.09 }$ & $ 65.16 \pm 0.68 $ & $ 2.0 \pm 0.1 $ & 470.7 / 462.0 & 0.38 & 5.9 \\
205.86 - 206.88 & ${ 4.76 }_{- 0.98 }^{+ 0.91 }$ & ${ -0.67 }_{- 0.08 }^{+ 0.05 }$ & ${ 81.74 }_{- 25.8 }^{+ 20.2 }$ & ${ -1.33 }_{- 0.24 }^{+ 0.13 }$ & ${ 261.8 }_{- 30.2 }^{+ 25.5 }$ & ${ -2.44 }_{- 0.12 }^{+ 0.05 }$ & $ 54.08 \pm 0.64 $ & $ 1.5 \pm 0.1 $ & 472.2 / 462.0 & 0.36 & 2.4 \\
206.88 - 207.91 & $0$ & $0$ & $0$ & $0$ & $0$ & $0$ & $ 0 $ & $ 0 $ & 0 & 0 & 0 \\
\hline
\end{tabular}
\end{table*}

\begin{table*}
\centering
\caption{Best fit parameters for the SBPL+BB model for the main emission episode of GRB160625B.}
\label{tab:sbplBB}
\scriptsize
\begin{tabular}{cccccccccccc}
\hline\hline
   \multicolumn{1}{c}{Time Bin} &
   \multicolumn{1}{c}{Norm.} &
   \multicolumn{1}{c}{$\alpha$} &
   \multicolumn{1}{c}{$E_{peak}$} &
   \multicolumn{1}{c}{$\beta$} &
   \multicolumn{1}{c}{BB Norm. ($10^{-5}$)} &
   \multicolumn{1}{c}{kT} &
   \multicolumn{1}{c}{Photon Flux} &
   \multicolumn{1}{c}{Energy Flux($10^{-5}$)} &
   \multicolumn{1}{c}{$\chi^{2} / DOF$} &
   \multicolumn{1}{c}{Prob} \\
   $[s]$ &  &  & $[keV]$ & & $[ph\, s^{-1} \,cm^{-2} \, keV^{-1}]$ & $[keV]$ &  $[ph \, s^{-1} \, cm^{-2}]$ & $[erg \, s^{-1} \, cm^{-2}]$ &  &   \\
 \hline

186.40 - 187.43 & ${ 4.67 }_{- 0.69 }^{+ 0.72 }$ & ${ -1.11 }_{- 0.04 }^{+ 0.03 }$ & ${ 3462.0 }_{- 333.0 }^{+ 609.0 }$ & ${ -2.46 }_{- 0.31 }^{+ 0.2 }$ & ${ 0.6 }_{- 0.2 }^{+ 0.2 }$ & ${ 66.3 }_{- 6.88 }^{+ 8.68 }$ & $ 20.53 \pm 0.44 $ & $ 2.4 \pm 0.1 $ & 449.2 / 462 & 0.66 \\
187.43 - 188.45 & ${ 5.6 }_{- 0.68 }^{+ 0.68 }$ & ${ -0.83 }_{- 0.03 }^{+ 0.03 }$ & ${ 1450.0 }_{- 95.5 }^{+ 152.0 }$ & ${ -2.09 }_{- 0.04 }^{+ 0.03 }$ & ${ 7.5 }_{- 1.2 }^{+ 1.7 }$ & ${ 43.7 }_{- 3.33 }^{+ 3.07 }$ & $ 79.61 \pm 0.69 $ & $ 8.0 \pm 0.2 $ & 523.0 / 462 & 0.03 \\
188.45 - 189.47 & ${ 15.07 }_{- 0.94 }^{+ 0.97 }$ & ${ -0.84 }_{- 0.02 }^{+ 0.02 }$ & ${ 997.2 }_{- 27.1 }^{+ 30.1 }$ & ${ -2.28 }_{- 0.02 }^{+ 0.02 }$ & ${ 25.0 }_{- 1.9 }^{+ 2.0 }$ & ${ 44.44 }_{- 1.27 }^{+ 1.29 }$ & $ 213.39 \pm 1.1 $ & $ 16.1 \pm 0.2 $ & 567.1 / 462 & $5.8 \cdot 10^{-4}$ \\
189.47 - 190.50 & ${ 15.5 }_{- 1.15 }^{+ 1.1 }$ & ${ -0.86 }_{- 0.02 }^{+ 0.02 }$ & ${ 747.2 }_{- 24.8 }^{+ 25.6 }$ & ${ -2.31 }_{- 0.03 }^{+ 0.02 }$ & ${ 33 }_{- 2.9 }^{+ 3.5 }$ & ${ 37.01 }_{- 1.35 }^{+ 1.25 }$ & $ 174.25 \pm 1.0 $ & $ 10.4 \pm 0.2 $ & 547.1 / 462 & $ 3.8 \cdot 10^{-3}$ \\
190.50 - 191.52 & ${ 12.86 }_{- 1.28 }^{+ 1.28 }$ & ${ -0.93 }_{- 0.03 }^{+ 0.02 }$ & ${ 538.4 }_{- 29.8 }^{+ 29.6 }$ & ${ -2.42 }_{- 0.07 }^{+ 0.03 }$ & ${ 29 }_{- 3.6 }^{+ 4.0 }$ & ${ 32.04 }_{- 1.52 }^{+ 1.48 }$ & $ 99.34 \pm 0.79 $ & $ 4.1 \pm 0.1 $ & 458.9 / 462 & 0.53 \\
191.52 - 192.55 & ${ 17.09 }_{- 2.4 }^{+ 1.95 }$ & ${ -1.04 }_{- 0.05 }^{+ 0.03 }$ & ${ 484.0 }_{- 52.0 }^{+ 42.0 }$ & ${ -2.47 }_{- 0.12 }^{+ 0.05 }$ & ${ 26.0 }_{- 3.3 }^{+ 3.7 }$ & ${ 31.19 }_{- 1.84 }^{+ 1.51 }$ & $ 78.58 \pm 0.73 $ & $ 2.6 \pm 0.1 $ & 434.7 / 462 & 0.81 \\
192.55 - 193.57 & ${ 12.08 }_{- 1.71 }^{+ 1.35 }$ & ${ -0.9 }_{- 0.05 }^{+ 0.03 }$ & ${ 388.2 }_{- 34.9 }^{+ 25.5 }$ & ${ -2.46 }_{- 0.07 }^{+ 0.04 }$ & ${ 30.0 }_{- 4.2 }^{+ 5.5 }$ & ${ 30.05 }_{- 2.27 }^{+ 1.68 }$ & $ 93.47 \pm 0.78 $ & $ 3.2 \pm 0.1 $ & 459.9 / 462 & 0.52 \\
193.57 - 194.59 & ${ 13.52 }_{- 1.18 }^{+ 1.02 }$ & ${ -0.9 }_{- 0.02 }^{+ 0.02 }$ & ${ 553.9 }_{- 21.4 }^{+ 17.1 }$ & ${ -2.59 }_{- 0.06 }^{+ 0.03 }$ & ${ 30.0 }_{- 3.1 }^{+ 3.8 }$ & ${ 34.22 }_{- 1.44 }^{+ 1.23 }$ & $ 123.53 \pm 0.87 $ & $ 5.1 \pm 0.1 $ & 456.5 / 462 & 0.56 \\
194.59 - 195.62 & ${ 15.25 }_{- 1.25 }^{+ 1.07 }$ & ${ -0.9 }_{- 0.02 }^{+ 0.02 }$ & ${ 540.6 }_{- 20.1 }^{+ 15.8 }$ & ${ -2.6 }_{- 0.06 }^{+ 0.03 }$ & ${ 34.0 }_{- 3.3 }^{+ 4.0 }$ & ${ 33.82 }_{- 1.32 }^{+ 1.13 }$ & $ 136.01 \pm 0.91 $ & $ 5.5 \pm 0.1 $ & 488.6 / 462 & 0.19 \\
195.62 - 196.64 & ${ 13.63 }_{- 1.27 }^{+ 1.06 }$ & ${ -0.9 }_{- 0.02 }^{+ 0.02 }$ & ${ 513.0 }_{- 22.3 }^{+ 16.3 }$ & ${ -2.64 }_{- 0.07 }^{+ 0.03 }$ & ${ 31.0 }_{- 3.3 }^{+ 4.1 }$ & ${ 33.34 }_{- 1.49 }^{+ 1.24 }$ & $ 118.0 \pm 0.85 $ & $ 4.5 \pm 0.1 $ & 484.5 / 462 & 0.23 \\
196.64 - 197.67 & ${ 11.59 }_{- 1.65 }^{+ 1.14 }$ & ${ -0.9 }_{- 0.04 }^{+ 0.02 }$ & ${ 451.6 }_{- 34.5 }^{+ 20.5 }$ & ${ -2.59 }_{- 0.1 }^{+ 0.04 }$ & ${ 36.0 }_{- 5.8 }^{+ 9.7 }$ & ${ 27.5 }_{- 2.33 }^{+ 1.59 }$ & $ 94.58 \pm 0.8 $ & $ 3.4 \pm 0.1 $ & 476.7 / 462 & 0.31 \\
197.67 - 198.69 & ${ 13.1 }_{- 1.31 }^{+ 1.09 }$ & ${ -0.92 }_{- 0.03 }^{+ 0.02 }$ & ${ 526.7 }_{- 25.7 }^{+ 20.2 }$ & ${ -2.53 }_{- 0.07 }^{+ 0.03 }$ & ${ 26.0 }_{- 3.6 }^{+ 4.7 }$ & ${ 31.44 }_{- 1.84 }^{+ 1.53 }$ & $ 99.99 \pm 0.81 $ & $ 4.1 \pm 0.1 $ & 461.0 / 462 & 0.5 \\
198.69 - 199.71 & ${ 11.44 }_{- 1.04 }^{+ 1.03 }$ & ${ -0.89 }_{- 0.02 }^{+ 0.02 }$ & ${ 536.9 }_{- 20.8 }^{+ 19.2 }$ & ${ -2.73 }_{- 0.08 }^{+ 0.05 }$ & ${ 25.0 }_{- 3.4 }^{+ 4.0 }$ & ${ 32.97 }_{- 1.72 }^{+ 1.71 }$ & $ 103.42 \pm 0.81 $ & $ 4.1 \pm 0.1 $ & 487.6 / 462 & 0.2 \\
199.71 - 200.74 & ${ 12.47 }_{- 1.06 }^{+ 1.03 }$ & ${ -0.84 }_{- 0.02 }^{+ 0.02 }$ & ${ 683.1 }_{- 25.4 }^{+ 24.8 }$ & ${ -2.31 }_{- 0.03 }^{+ 0.02 }$ & ${ 22.0 }_{- 2.9 }^{+ 3.7 }$ & ${ 36.65 }_{- 2.1 }^{+ 1.99 }$ & $ 145.72 \pm 0.93 $ & $ 8.8 \pm 0.2 $ & 518.8 / 462 & 0.03 \\
200.74 - 201.76 & ${ 12.8 }_{- 1.08 }^{+ 1.04 }$ & ${ -0.85 }_{- 0.02 }^{+ 0.02 }$ & ${ 586.9 }_{- 23.6 }^{+ 22.7 }$ & ${ -2.33 }_{- 0.03 }^{+ 0.02 }$ & ${ 25.0 }_{- 3.1 }^{+ 3.7 }$ & ${ 34.79 }_{- 1.73 }^{+ 1.63 }$ & $ 134.58 \pm 0.91 $ & $ 7.0 \pm 0.1 $ & 479.8 / 462 & 0.27 \\
201.76 - 202.79 & ${ 16.79 }_{- 1.37 }^{+ 1.16 }$ & ${ -0.95 }_{- 0.02 }^{+ 0.02 }$ & ${ 562.9 }_{- 21.1 }^{+ 15.1 }$ & ${ -2.8 }_{- 0.09 }^{+ 0.04 }$ & ${ 26.0 }_{- 2.9 }^{+ 3.7 }$ & ${ 34.25 }_{- 1.6 }^{+ 1.36 }$ & $ 118.77 \pm 0.86 $ & $ 4.4 \pm 0.1 $ & 433.6 / 462 & 0.82 \\
202.79 - 203.81 & ${ 17.12 }_{- 1.9 }^{+ 1.36 }$ & ${ -1.01 }_{- 0.03 }^{+ 0.02 }$ & ${ 470.8 }_{- 34.5 }^{+ 21.6 }$ & ${ -2.63 }_{- 0.13 }^{+ 0.04 }$ & ${ 15 }_{- 2.5 }^{+ 3.4 }$ & ${ 33.31 }_{- 2.65 }^{+ 1.92 }$ & $ 83.83 \pm 0.75 $ & $ 2.8 \pm 0.1 $ & 489.0 / 462 & 0.19 \\
203.81 - 204.83 & ${ 12.32 }_{- 1.76 }^{+ 1.21 }$ & ${ -0.98 }_{- 0.04 }^{+ 0.02 }$ & ${ 490.3 }_{- 44.7 }^{+ 30.5 }$ & ${ -2.46 }_{- 0.11 }^{+ 0.04 }$ & ${ 15.0 }_{- 3.3 }^{+ 5.1 }$ & ${ 30.16 }_{- 3.36 }^{+ 2.36 }$ & $ 67.95 \pm 0.69 $ & $ 2.6 \pm 0.1 $ & 473.9 / 462 & 0.34 \\
204.83 - 205.86 & ${ 12.75 }_{- 1.66 }^{+ 1.19 }$ & ${ -1.0 }_{- 0.04 }^{+ 0.02 }$ & ${ 458.7 }_{- 38.2 }^{+ 22.8 }$ & ${ -2.7 }_{- 0.17 }^{+ 0.06 }$ & ${ 17.0 }_{- 2.7 }^{+ 3.5 }$ & ${ 31.93 }_{- 2.38 }^{+ 1.75 }$ & $ 65.37 \pm 0.68 $ & $ 2.1 \pm 0.2 $ & 472.2 / 462 & 0.36 \\
205.86 - 206.88 & ${ 8.45 }_{- 2.13 }^{+ 1.48 }$ & ${ -0.91 }_{- 0.1 }^{+ 0.05 }$ & ${ 288.4 }_{- 45.7 }^{+ 33.6 }$ & ${ -2.43 }_{- 0.12 }^{+ 0.04 }$ & ${ 23.0 }_{- 6.3 }^{+ 10.7 }$ & ${ 25.15 }_{- 4.92 }^{+ 2.63 }$ & $ 54.18 \pm 0.64 $ & $ 1.5 \pm 0.2 $ & 471.0 / 462 & 0.38 \\
206.88 - 207.91 & $0$ & $0$ & $0$ & $0$ & $0$ & $0$ & $ 0$ & $ 0 $ & 0 & 0 \\
\hline
\end{tabular}
\end{table*}

\end{appendix}

\end{document}